\documentclass{article}

\usepackage{ulem}
\usepackage{makecell}    
\usepackage{color}

\usepackage[utf8]{inputenc}
\usepackage{authblk}
\usepackage{setspace}
\usepackage[margin=1.25in]{geometry}
\usepackage{graphicx}
\graphicspath{ {./figures/} }
\usepackage{subcaption}
\usepackage{amsmath}
\usepackage{amssymb}

\usepackage[style=nejm, 
citestyle=numeric-comp,
sorting=none,
backend=bibtex]{biblatex}
\title{Chiral high harmonic generation via coherent superposition of multi-channel interband transitions}

\author[1,2,$\dag$]{Ruoxuan~Ding}
\author[3,4,*,$\dag$]{Chen~Qian}
\author[3,*]{Zhilin~Li}
\author[1]{Tianchen~Hu}
\author[2]{Chunmei~Zhang}
\author[5,1]{Tao~Dong}
\author[2]{Dong~Wu}
\author[2,*]{Li~Yue}
\author[5,1,2,*]{Nan-Lin~Wang}

\affil[1]{International Center for Quantum Materials, School of Physics, Peking University, Beijing 100871, China.}
\affil[2]{Beijing Academy of Quantum Information Sciences, Beijing 100913, China.}
\affil[3]{Beijing National Laboratory for Condensed Matter Physics, Institute of Physics, Chinese Academy of Sciences, Beijing 100190, China.}
\affil[4]{Institute of Ultrafast Optical Physics, Department of Applied Physics \& MIIT Key Laboratory of Semiconductor Microstructure and Quantum Sensing, Nanjing University of Science and Technology, Nanjing 210094, China.}
\affil[5]{Tsung-Dao Lee Institute, Shanghai Jiao Tong University, Shanghai 200240, China.}
\affil[*]{Address correspondence to: cqian@iphy.ac.cn, lizhilin@iphy.ac.cn, yueli@baqis.ac.cn, nlwang@sjtu.edu.cn}
\affil[$\dag$]{These authors contributed equally to this work.}

\date{}

\begin{document}
\maketitle

\begin{abstract}
Circularly polarized light, with its specific handedness, is a valuable tool for investigating  the chirality related properties of materials, including their structural, electronic, magnetic and topological characteristics. This has led to efforts in creating wavelength-tunable coherent light with circular or elliptical polarization through a sub-cycle ultrafast process high-harmonic generation (HHG). Existing approaches to chiral HHG, however, typically rely on multidimensional driving fields, materials engineering or complex optical architectures. 
Here we uncover a distinct mechanism without such complexity, where a single linearly polarized driving field generates chiral HHG from a pristine crystal. We show that, in the Weyl semimetal NbP, one linearly polarized near-infrared driving field generates high harmonics with pronounced ellipticity.
Semi-conductor Bloch equations (SBEs) reveal that this ellipticity originates from the coherent superposition of multiple interband transitions, each contributing harmonics with distinct linear polarizations and relative phases. We further substantiate this mechanism by shifting the driving photon energy into the mid-infrared, which strongly suppresses the ellipticity as the number of accessible interband transition channels is reduced. These results establish a simple and general mechanism for chiral HHG, and show that quantum materials can transform an apparently achiral driving field into polarization-tunable ultrafast chiral emission.
\end{abstract}


\section{Introduction}

 Chirality, a fundamental geometric property, has profound implications across the physical, chemical and biological sciences. 
Yet, identifying chirality is challenging due to the identical nature of objects with opposite handedness, where all scalar physical properties are the same. 
They become distinguishable only through interactions with other chiral entities. Elliptically and circularly polarized light is itself chiral and is therefore essential for distinguishing structural and magnetic chirality\cite{MunLight2020}, probing chiral dynamics in particles\cite{ChanPRL2016,HanNature2025} and collective excitations\cite{wang2024ChiralPhononReview}, and engineering material functionalities accordingly\cite{Kim2021Spintronics,LuoScience2023}. While high-harmonic generation (HHG) has emerged as a key route to tunable broadband ultrafast light sources extending into the extreme-ultraviolet and X-ray regimes, a straightforward approach to chiral HHG is highly desirable. From a symmetry perspective, however, elliptically polarized harmonics cannot be generated from a single linearly polarized driving field in isotropic gas targets\cite{PhysRevA.49.2117,DietrichPRA1994}, which remain the standard medium for HHG experiments. As a result, existing approaches to chiral HHG are typically complex, requiring either tailored two-dimensional driving fields, such as bichromatic counter-rotating laser fields\cite{kfir2015circularEUV,Fan2015GenerationEUV}, or molecular pre-alignment induced by an additional laser pulse\cite{Skantzakis2016GenerationEUVthroughPreAlinedMolecule,2009CO2linear2ellipCalculation,2010PrlN2linear2ellipExp}.

Crystalline systems, by contrast, possess intrinsic anisotropy arising from their periodic atomic arrangement, offering rich opportunities to control the polarization of HHG. To generate chiral HHG using a linearly polarized driver, one promising strategy is therefore to exploit crystalline targets. In strained graphene, elliptically polarized high-order harmonics can arise from the intricate interplay between interband and intraband electron dynamics\cite{RanaPRB2024}. In twisted hBN flakes, tailored phase matching enables purely circularly polarized HHG emission regardless of the incident polarization\cite{Ma2025prlVdwPolarizationControl}. 
In engineered photonic structures, a related mechanism has recently been demonstrated in achiral resonant metasurfaces, where chiral third-harmonic generation (THG) arises from the coherent superposition of several resonant modes and the uncoupled third-harmonic field\cite{Lai2025}. 
Nevertheless, 
generating chiral HHG simply in pristine solids with a single linearly polarized driver remains challenging, despite being the most straightforward route. 
Theoretical studies have proposed several possible mechanisms, including the averaging of interband transitions over different \textbf{k} points\cite{Dong2021TheoryLinear2EllipticalGraphene,Zurron2019TheoryLinear2EllipticalGraphene} and Berry-curvature effects in topological phases\cite{silva2019topological}. 
However, a general feasible mechanism, validated by experimental realization, remains elusive.

In this work, we elucidate a general mechanism of chiral HHG originating from the band structure of a Weyl semimetal (WSM). We investigate the HHG response of NbP, a member of the TaAs family of WSMs. This family exhibits both non-trivial topological characteristics\cite{xu2015TaAsFermiArc,lv2015TaAsfermiArc,lv2015TaAsWeylNode,zhang2016TaAsNegativeLMR,huang2015TaAsNegativeLMR,2015NpNbpMagnetoresistance} and pronounced nonlinear optical responses\cite{wu2017TaAsSHG,ma2019nonlinear,osterhoudt2019colossal,ma2017TaAsPumpCPGE}. Previous HHG studies on WSMs have primarily focused on the influence of topology on the harmonic response\cite{2022PrbWeyl2HHGtheory,2022PraHHGin3dWeylTheory,2022PrbWeylAHETheory,2024PrbWeylMergingWeylNodeHHGTheory,2024AplWeylHHGIntensityTuningTheory,2023PrbRoleOfTopoChargeWeylHHGTheory}. Here, instead, we take advantage of a favourable feature of Weyl semimetals: their gapless and complex band structure supports multiple interband transitions even at low harmonic orders, providing an attractive platform for investigating multichannel coherent superposition, which is considered crucial in non-perturbative nonlinear processes\cite{DirectIndirectInterfereHHG2015,Linear2ellipticalHHGGaAs2023,HSGinterferenceBloch2017,MoS2Multiband2022,luu2016SBEsQuantumInterference,NPho2014CEPstableTHzQuantumInterference}.
 We find that the coherent superposition of interband channels driven by a single linearly polarized near-infrared laser leads to pronounced ellipticity in low-order harmonics. Numerical simulations using Semi-conductor Bloch equations (SBEs) further corroborate the experimental observations. Guided by the theoretical analysis, we further examine the ellipticity of HHG driven by a 4,000$\mathrm{nm}$ pulse, for which the ellipticity is strongly suppressed as
the number of accessible interband transition channels is reduced, thereby providing additional support for multichannel superposition as the origin of the chiral HHG emission. Microscopically, this superposition arises from the coherent evolution and interband recollision of carriers on femtosecond and sub-cycle timescales.

\section{Results}
\begin{figure}[htbp]
    \begin{center}
	   \includegraphics[page=1, width=1\textwidth, trim=0in 0in 0in 0in, clip]{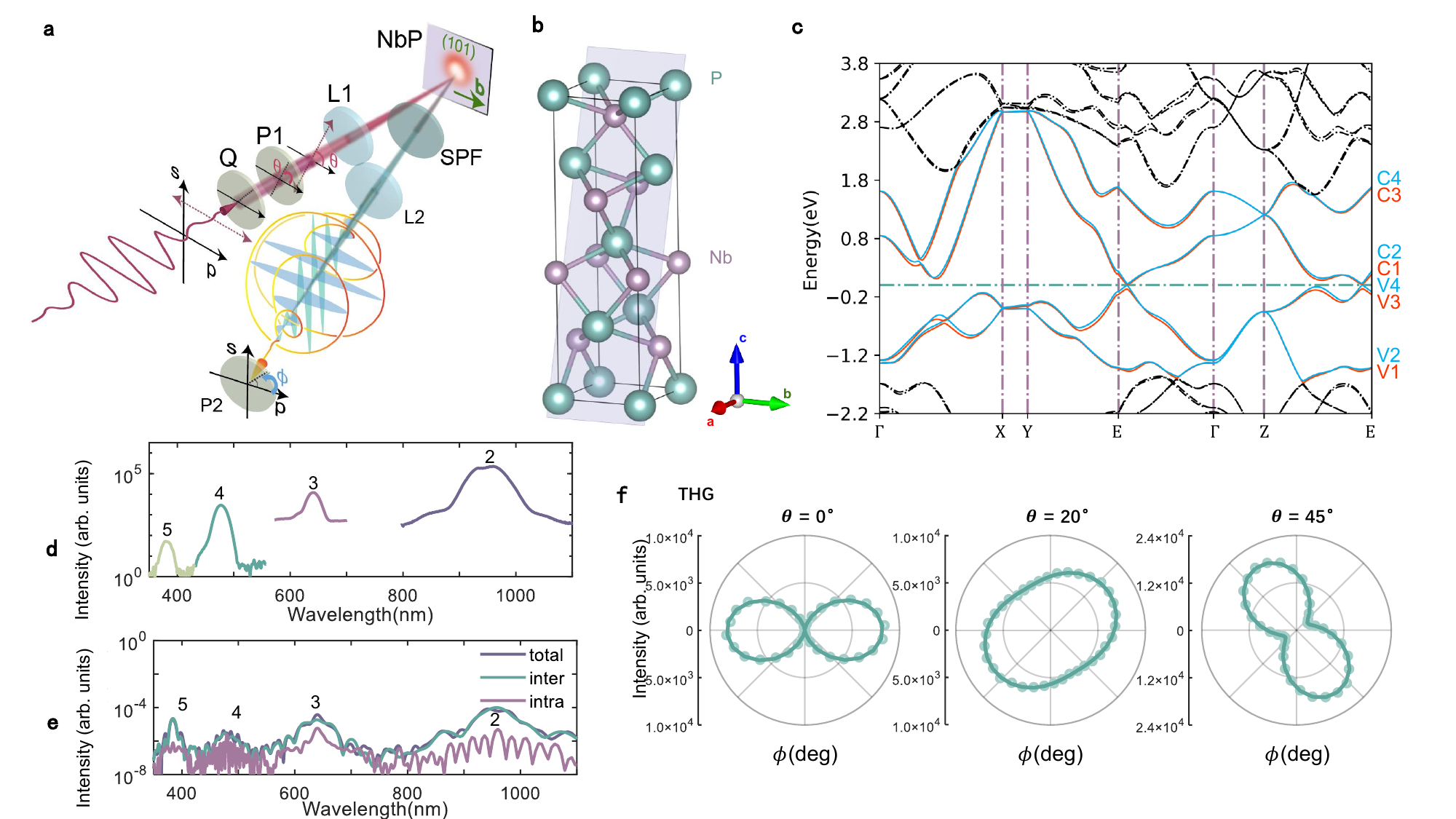}
\caption{\textbf{Structure and chiral harmonic generation in NbP.} \textbf{(a)}, Schematic of the optical setup. Measurements were performed in a reflection geometry using a 1,920$\mathrm{nm}$ driving laser generated by an optical parametric amplifier (OPA), and the emitted harmonics were collected by a commercial spectrometer. Q, quarter-wave plate with its fast axis oriented at 45$^\circ$ to the horizontal direction; P1, polarizer, with $\theta$ denoting the angle between its transmission axis and the \textit{b} axis; P2, polarizer, with $\phi$ denoting the angle between its transmission axis and the \textit{b} axis; L1 and L2, focusing lenses; SPF, short-pass filters. \textbf{(b)}, Crystal structure of NbP in a body-centred tetragonal lattice. The shaded plane indicates the (101) surface. \textbf{(c)}, Electronic band structure along high-symmetry directions with SOC included. The red and blue solid lines indicate the eight bands used in the SBEs. These bands are labelled C1--C4 for the conduction bands and V1--V4 for the valence bands, ordered in energy from low to high. \textbf{(d)}, Harmonic spectrum emitted from the NbP crystal on the (101) surface, showing harmonics up to the fifth order (380 $\mathrm{nm}$) at $\theta = 5^\circ$. \textbf{(e)}, Calculated interband and intraband contributions to chiral harmonic generation in NbP, showing the dominance of the interband contribution. The calculated spectrum is truncated to the same wavelength window as the experiment for direct comparison. 
 \textbf{(f)}, Polarization state of the third-harmonic emission ($I = 2.7\times10^{11}\,\mathrm{W\,cm^{-2}}$), showing nearly circularly polarized emission at $\theta = 20^\circ$ (see main text for definition of ellipticity).}
	   \label{fig:1}
    \end{center}
\end{figure}

NbP crystallizes in a non-centrosymmetric tetragonal structure (space group I4$_1$md), as shown in Fig.~\ref{fig:1}b. Along [001], the tetragonal structure has a fourfold screw axis; consequently, its Brillouin zone exhibits a C$_4$ rotational symmetry about $k_z$ (Supplementary Section S2).
NbP offers several advantages for investigating high-harmonic generation (HHG). First, it exhibits a strong nonlinear optical response: the third-order nonlinearity of NbP films is approximately one order of magnitude larger than that of Ge films\cite{tilmann2022NbPFilm3HG}. Second, its electronic structure is particularly favourable, featuring linearly dispersing Weyl cones and multiple bands near the Fermi level separated by small direct gaps. The former supports intense harmonic emission of various orders, whereas the latter suggests that multiple interband transition channels can contribute even to low-order harmonics. Fig.~\ref{fig:1}c shows the calculated band structure along high-symmetry directions in the Brillouin zone with spin–orbit coupling (SOC) included. A pronounced Weyl node — the hallmark of a Weyl semimetal — appears near a high-symmetry point and exhibits the expected linear conical dispersion. To generate HHG, we used a linearly polarized excitation pulse at 1,920 $\mathrm{nm}$ with a duration of 60 fs. Measurements were performed in reflection geometry on the [101] crystal face. As shown in Fig.~\ref{fig:1}d, both even- and odd-order harmonics up to the fifth order were observed. The microscopic origin of HHG in Weyl semimetals remains under active debate. In $\beta$-WP$_2$, HHG at 1,900 $\mathrm{nm}$ has been attributed to intraband Bloch oscillations and Berry-curvature effects\cite{lv2021WeylWP2HHG}, whereas interband transitions have also been proposed to play a crucial role in the same material\cite{PrbWP2Calculation}. In contrast, our SBEs indicate that, in NbP driven at 1,920 $\mathrm{nm}$, harmonics of different orders are all dominated predominantly by interband transitions (Fig.~\ref{fig:1}e). 
We further decomposed the calculated THG into intraband and interband contributions and examined their polarization states (Supplementary Section S12).
Additional evidence for the predominance of the interband mechanism is presented below.

\begin{figure}[h]
    \begin{center}
        \includegraphics[page=2, width=1\textwidth, trim=0in 0in 0in 0in, clip]{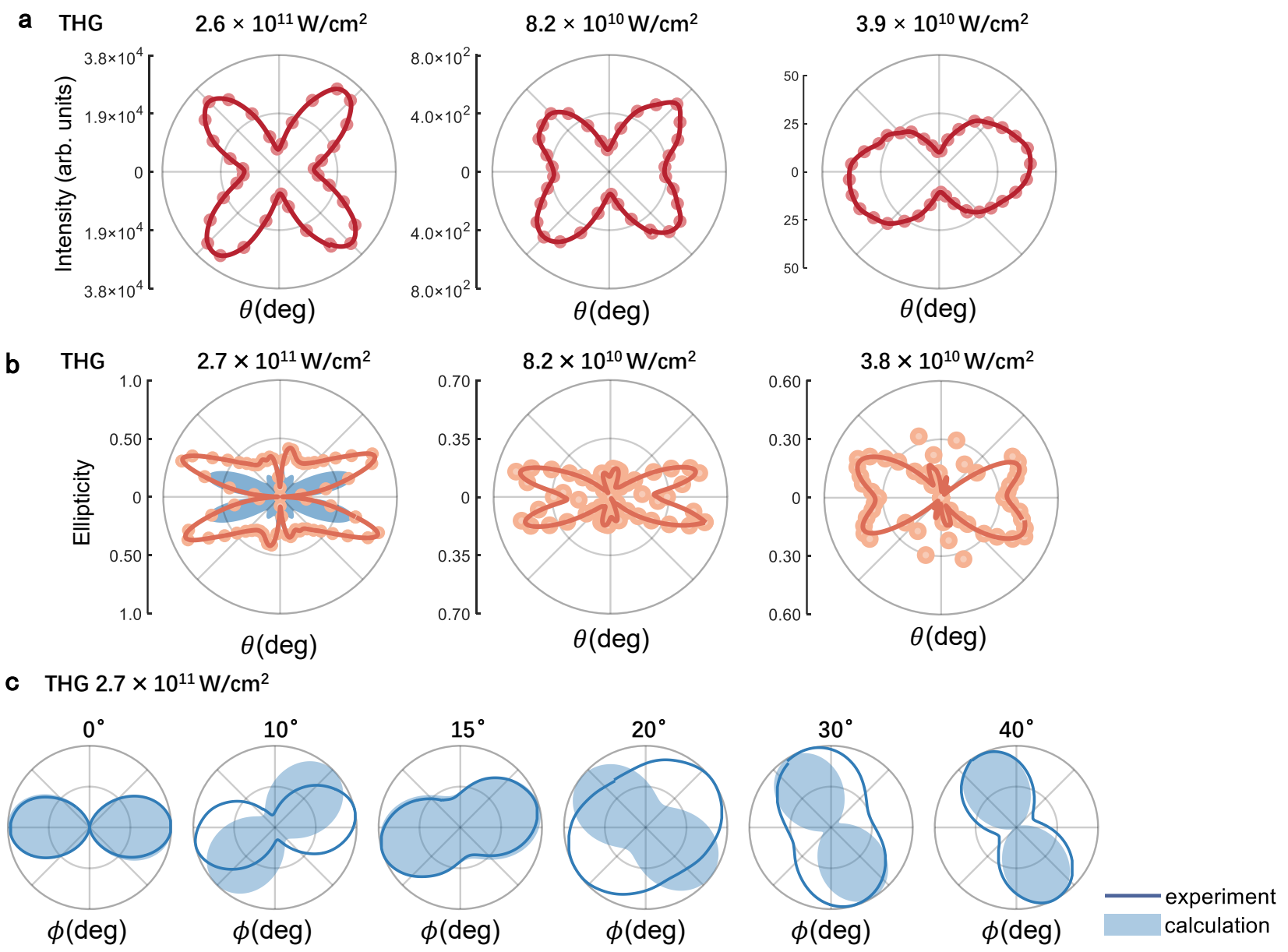}
\caption{\textbf{Orientation dependence and chirality of harmonic emission in NbP.} \textbf{(a)}, Orientation-dependent intensities of THG from the NbP (101) surface at different powers. Experimental data are shown as red circles, while lines are generated by interpolation solely as a visual guide for the eye. \textbf{(b)}, Experimentally measured $\epsilon$ as a function of $\theta$. Experimental data are shown as orange circles.
Measurements were performed only over a $180^{\circ}$ range; for display, the data are repeated every $180^{\circ}$ to show the full $360^{\circ}$  dependence.
all data points are single determinations, except for the two symmetric points with the largest $\epsilon$ values. These two points are plotted as the mean $\pm$ standard deviation (SD) of n = 3 independent measurements: $\epsilon$ = 0.90 ± 0.015;
$\theta$ is fixed by the experimental conditions; lines are generated by interpolation solely as a visual guide for the eye. The blue shaded region in the left panel indicates the SBEs-calculated $\epsilon$. 
 \textbf{(c)}, Polarization-resolved measurements ($\phi$ scans; blue solid lines) and corresponding calculations (light blue shading) of the emitted harmonic field for $\theta = 0^\circ$–$40^\circ$. The data in \textbf{(c)} are normalized. The measurements were performed at a peak intensity of $\sim 2.7\times10^{11}\,\mathrm{W\,cm^{-2}}$.}
	   \label{fig:2}
    \end{center}
\end{figure}

Having established that the calculated harmonic response is dominated by interband transitions, we next examined the polarization state of the emitted THG. 
To this end, the linear polarization of the driver was rotated, with $\theta$ defined as the angle between the driver polarization and the crystal \textit{b} axis, and $\phi$ defined as the angle between the transmission axis of the output analyzer (P2) and the \textit{b} axis. For each $\theta$, the harmonic intensity 
$I(\phi)$ was recorded by rotating P2 before the spectrometer.
A natural expectation for a single linearly polarized driving field is that the harmonic simply inherits the linear polarization of the driver. The representative analyzer scans in Fig.~\ref{fig:1}f, measured at pump intensity $I = 2.7\times10^{11}\,\mathrm{W\,cm^{-2}}$, show that this expectation fails in a nontrivial way. At $\theta = 0^{\circ}$, the THG is indeed nearly linearly polarized along the driver, consistent with the mirror symmetry. At $\theta = 20^{\circ}$, however, the harmonic becomes strongly elliptical. At $\theta = 45^{\circ}$ the ellipticity is markedly reduced again and its major axis is rotated substantially away from the driving polarization. The polarization response is therefore strongly nonmonotonic in $\theta$: leaving the high-symmetry direction first turns on a pronounced elliptical component, but a further rotation suppresses it. Such a nonmonotonic polarization response is incompatible with a simple projection of the linearly polarized driver onto the harmonic field. It indicates that the emitted polarization is governed by a more complex, orientation-dependent mechanism, which will be systematically investigated in the following sections. The complete $\theta$-dependent $\phi$ scans and their evolution with pump power are presented in Fig.~\ref{fig:2}.

Fig.~\ref{fig:2} summarizes the orientation- and power-dependent polarimetric response of THG.
Figure~\ref{fig:2}a shows the harmonic intensity as a function of the linear polarization angle of the driving field, denoted by $\theta$, with the polarizer P2 for HHG output removed. Intriguingly, the angular dependence changes dramatically with increasing excitation power (Fig.~\ref{fig:2}a). At low power, the pattern consists of two lobes aligned along the \textit{b} axis ($\theta = 0^\circ$). As the power increases, it evolves into a butterfly-shaped profile and ultimately into a fourfold-symmetric X-shaped pattern.
{We also performed the same orientation-dependent intensity measurements for the fourth and fifth harmonics (Supplementary Section S11).}
﻿{The intensity anisotropy and ellipticity of THG in Fig.~\ref{fig:2} are closely connected. As the pump intensity increases, the two-lobe pattern evolves through a butterfly-shaped profile into four lobes, accompanied by a relative suppression of emission near $\theta = 0^{\circ}$ (Fig.~\ref{fig:2}a). Over the same intensity range, the THG yield at $\theta = 15^{\circ}$ exhibits a pronounced kink, unlike the smoother nonperturbative dependence at $\theta = 45^{\circ}$ (Supplementary Fig.~\ref{fig:S14}a).
{
 This kink implies that the system enters the non-perturbative regime, in which the 
$n$-th harmonic yield deviates from the perturbative power law $I_n \propto I_{\mathrm{pump}}^n$. Because the petal pattern changes in different pump-power ranges, the pump-intensity dependence of the harmonic yield may also differ from the usual case. 
Concurrently,}
 enhanced ellipticity emerges at nearby off-symmetry orientations (Fig.~\ref{fig:2}b). Channel-resolved SBE calculations further clarify this connection. Upon coherently adding interband contributions in descending order of their individual THG intensities, the cumulative major-axis power exhibits a pronounced dip at $\theta = 15^{\circ}$, but increases overall at $\theta = 45^{\circ}$, whereas the minor-axis powers remain of the same order of magnitude (Supplementary Fig.~\ref{fig:S14}b and c). Such destructive interference suppresses the dominant polarization component, favoring a more balanced polarization ellipse and enhanced ellipticity. Crucially, mirror symmetry enforces linear harmonic emission at $\theta = 0^{\circ}$, whereas rotation away from this direction allows a transverse response whose amplitude and phase are governed by multichannel coherence.}
We define the ellipticity as $\epsilon = \frac{E_1}{E_0}$, where $E_0$ and $E_1$ denote the electric-field amplitudes along the major and minor axes of the harmonic polarization ellipse, respectively, and $\epsilon$ ranges from 0 for linear polarization to 1 for circular polarization. 
{Experimentally, we measured the harmonic intensity $I(\phi)$. $I_0$ and $I_1$ denote the fitted intensities along the major and minor axes of the harmonic polarization ellipse, respectively. The corresponding field amplitudes are $E_{0,1} = \sqrt{I_{0,1}}$.}
Fig.~\ref{fig:2}b summarizes the ellipticity of THG extracted from systematic polarization-resolved measurements as a function of $\theta$ at the three pump intensities corresponding to Fig.~\ref{fig:2}a. When $\theta = 0^\circ$ or $90^\circ$, the harmonic output is linearly polarized along the incident laser polarization. This behaviour is enforced by crystal symmetry, specifically by the mirror plane perpendicular to the \textit{b} axis. For lower-symmetry directions, however, these symmetry constraints are lifted, allowing elliptically polarized HHG to emerge, and the major axis of the harmonic polarization is markedly rotated with respect to the polarization direction of the pump. {As the pump intensity increases, the maximum ellipticity grows, showing that the chiral response becomes more pronounced in the non-perturbative regime.} We have also checked the left or right handedness of the chiral harmonics (Supplementary Section S7). 
﻿
In GaAs, harmonic ellipticity has been reported, which is the only reported case of chiral HHG generated by a single linearly polarized driver in a pristine crystal. However, its chiral HHG is largely confined to driving field strength within the crossover region from perturbative to non-perturbative regime\cite{2022SekiguchiPrbGaAsLinear2Elliptical}. There, the ellipticity is attributed to interference between emission channels of different nonlinear orders, which dominate the perturbative and extreme-nonlinearity regimes, respectively. By contrast, in NbP the ellipticity becomes more pronounced in the non-perturbative regime at high pump intensity. This substantial ellipticity in the nonperturbative regime points to a distinct mechanism for chiral harmonic generation.

We next examine the polarimetric properties of THG as a function of the orientation of the linearly polarized driving field (Fig.~\ref{fig:2}c), from which two key conclusions emerge. First, $\epsilon$ increases as the driving polarization is rotated away from high-symmetry crystalline directions. The highest measured $\epsilon$ occurs at $\theta = 20^{\circ}$ and $25^{\circ}$, where $\epsilon$ reaches 0.83 and 0.87. Second, the major axis of the polarization ellipse is initially nearly parallel to the driving-field orientation, but, after the ellipticity reaches its maximum, it rotates to become nearly perpendicular to the driver. This behaviour suggests the presence of multiple quantum-transition pathways with different phases and comparable amplitudes in the angular range where the ellipticity is maximal, enabling their coherent superposition. These polarization properties are reproduced by SBEs (the shaded region in Fig.~\ref{fig:2}c), although 
there exist certain discrepancies in polarization direction between theory and experiment (which will be explained later).
Here, we only focus on the third harmonic as the representative response because it provides the strongest and most reliable signal. By contrast, ellipticity in the fourth harmonic may further involve Berry-curvature-related contributions and therefore does not provide an unambiguous signature of the mechanism discussed here. Higher-order signals, including the fifth harmonic, are weaker and do not yield additional physical insight ({Supplementary Section S11}).

To identify the physical origin of the observed chiral harmonics in NbP, we first distinguish the roles of intraband acceleration and interband transitions. Intraband electron acceleration is not directly affected by dephasing, except for long-trajectory processes, and therefore preserves time-reversal symmetry. In a time-reversal-symmetric crystal, the band dispersion satisfies $\varepsilon(\mathbf{k}) = \varepsilon(-\mathbf{k})$. The group velocity associated with odd-harmonic emission, $\mathbf{v}(\mathbf{k}) = \nabla_{\mathbf{k}} \varepsilon(\mathbf{k})$, then obeys $\mathbf{v}(\mathbf{k}) = -\mathbf{v}(-\mathbf{k})$. Under linearly polarized excitation, the harmonic components perpendicular to the driving field at momenta $\mathbf{k}$ and $-\mathbf{k}$ are equal in magnitude but opposite in sign, and therefore cancel. As a result, the emitted harmonics remain strictly polarized along the driving field. Intraband dynamics alone thus cannot generate elliptically polarized high harmonics in a time-reversal-symmetric crystal.

In conventional semiconductors with large bandgaps, low-order harmonics are typically dominated by intraband currents, whereas interband contributions become more important only at higher orders. Elliptical low-order harmonics are therefore rarely expected. In the zero-gap Weyl semimetal NbP, however, the complex band structure near the Fermi level (Fig.~\ref{fig:1}c) allows multiple interband transition channels to contribute even to low-order harmonics, which is discussed in detail in Supplementary Section S3.
In addition, the observed even-order harmonics polarized parallel to the driving field cannot be attributed to intraband Berry-curvature effects (Supplementary Section S9). Together, these observations support the interpretation that interband transitions underlie the elliptically polarized odd-order HHG in NbP.

\begin{figure} [h]
	\begin{center}
		\includegraphics[page=3, width=1\textwidth, trim=0in 0in 0in 0in, clip]{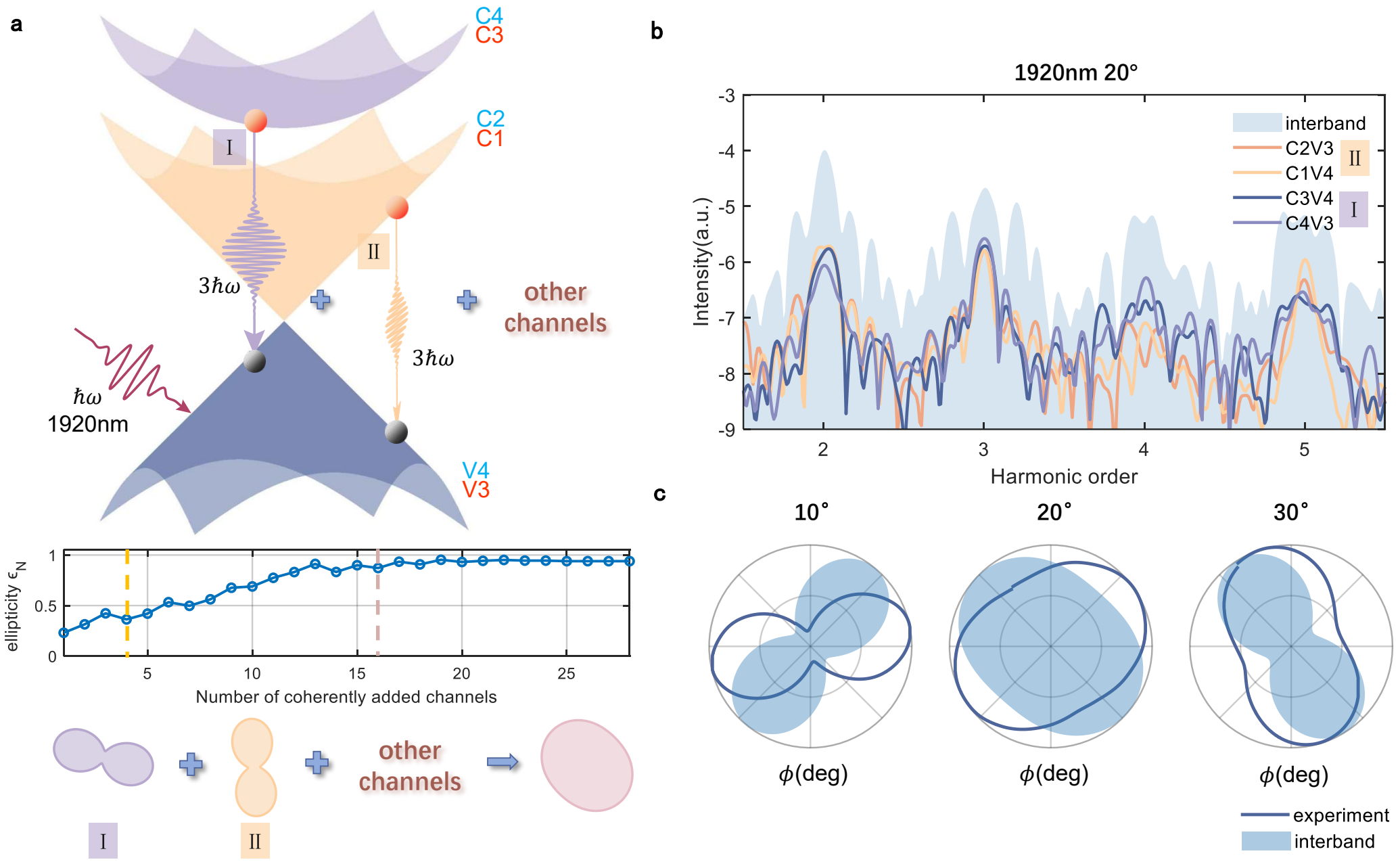}
		\caption{\textbf{Mechanism of chiral harmonics in NbP.} \textbf{(a)}, Schematic illustration of the emergence of elliptical polarization through the coherent superposition of multiple linearly polarized contributions from different interband channels. 
The middle panel shows the calculated interband ellipticity versus number of coherently added channels. The yellow dashed line indicates the cumulative result including only groups I and II; the pink dashed line indicates the situation when the 16 valence-to-conduction-band transition channels are taken into account (i.e. groups I, II, and III). The ellipticity increases with the number of channels, approaching its saturated value once nearly all transitions between the valence and conduction bands have been considered. \textbf{(b)}, Calculated harmonic intensity from the four most intense interband channels (solid lines) and from the total interband response including all 28 channels (light blue shading) at a driving wavelength of 1,920 $\mathrm{nm}$. \textbf{(c)}, Calculated polarization of the third harmonic from the total interband response (28 channels) for $\theta = 10^\circ$–$30^\circ$ (blue shading), together with the experimental data (solid lines). The data in \textbf{(c)} are normalized.}
		\label{fig:3}
	\end{center}
\end{figure}

\begin{figure} [h]
	\begin{center}
		\includegraphics[page = 4, width=1\textwidth, trim=0in 0in 0in 0in, clip]{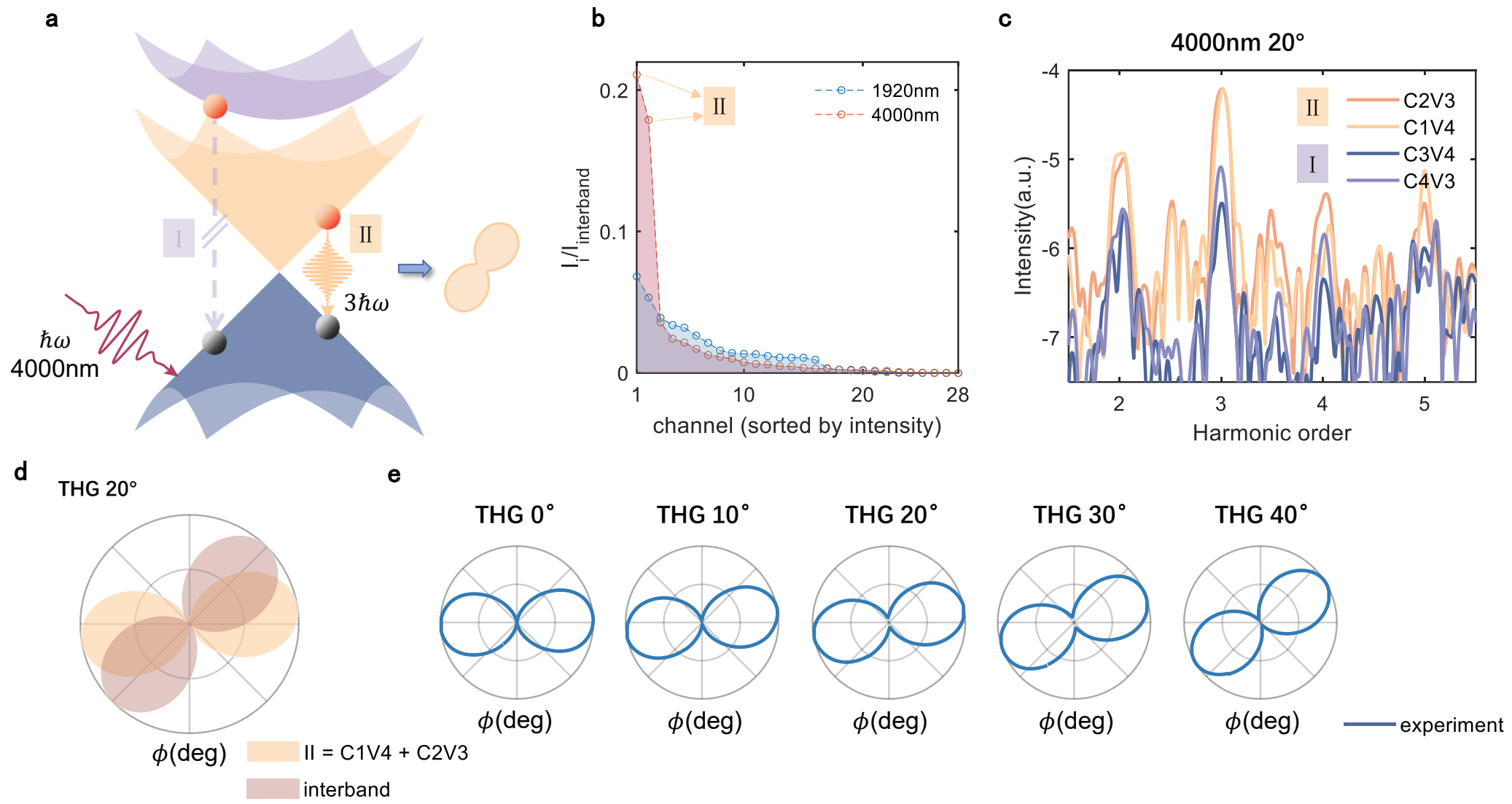} 
		\caption{\textbf{Wavelength dependence of chiral harmonic generation.} \textbf{(a)}, Schematic of third-harmonic generation driven at 4,000 $\mathrm{nm}$. The photon energy is too low to efficiently excite electrons into the C3 and C4 bands, rendering the C3V4 and C4V3 channels negligible and favouring linearly polarized harmonic emission. \textbf{(b)}, Intensity of the third harmonic generated by each channel, normalized to that of the total interband response (28 channels). The horizontal axis shows the channel index sorted by third-harmonic intensity. Blue and red circles denote the results for 1,920 $\mathrm{nm}$ and 4,000 $\mathrm{nm}$, respectively; the two largest red contributions correspond to C2V3 and C1V4. \textbf{(c)}, Calculated harmonic response from groups I and II. The contribution of group I to the third harmonic is strongly suppressed compared with that of group II. \textbf{(d)}, Polarization of the third harmonic from group II and from the total interband response (28 channels) at $\theta = 20^\circ$. The data are normalized. \textbf{(e)}, Measured polarization of the third harmonic for $\theta = 0^\circ$–$40^\circ$. The data in \textbf{(d)} and \textbf{(e)} are normalized.}
		\label{fig:4}
	\end{center}
\end{figure}

We next analyse the contribution of individual interband channels to elucidate the microscopic origin of the chiral response. In the eight-band model, there are 28 possible interband channels ($C_8^2 = 28$), whose combined contribution gives rise to strongly elliptical THG. By contrast, the response from any single channel is typically linear or only weakly elliptical (Supplementary Section S10). The pronounced ellipticity therefore emerges from the joint contribution of many channels with distinct polarization directions and relative phases. This condition is difficult to satisfy when only a few channels are involved. As illustrative examples, we consider the four channels with the highest intensities. The two channels in group I, corresponding to transitions between a Weyl valence band and a trivial conduction band (C4V3 and C3V4; see Supplementary Section S10 for the grouping criteria), combine to produce only weak ellipticity. The two channels in group II, corresponding to transitions between Weyl bands (C1V4 and C2V3), yield a similar outcome. Moreover, although the polarization directions of these two groups are nearly orthogonal, their relative phase is not close to $\pi/2$, and their combination still produces an almost linearly polarized output (Supplementary Section S10). 
We refer to all remaining valence-to-conduction channels as group III. Together, groups I, II and III comprise all valence-to-conduction channels. 
We find that the harmonic ellipticity remains small when only the four most intense channels (groups I and II) are retained (see Fig.~\ref{fig:3}a), and it grows substantially only as the larger set of non-negligible channels in group III is included, approaching its saturated value once nearly all transitions between the valence and conduction bands have been considered.
 The remaining 12 weaker contributions, which lie outside groups I, II and III and arise from transitions among valence bands and among conduction bands, induce only negligible changes in the ellipticity.
Importantly, no small subset of channels dominates the total interband response, as shown in Fig.~\ref{fig:4}b. Instead, 16 channels make non-negligible contributions, corresponding precisely to the 16 valence-to-conduction-band transitions.

The coexistence of multiple channels with comparable amplitudes, but different polarization directions and phases, favours the emergence of harmonics with pronounced ellipticity. This picture further implies that maximum ellipticity should not generally coincide with the orientation that maximizes the harmonic intensity (Fig.~\ref{fig:2}a). The latter is more likely to occur when several channels emit along similar polarization directions and therefore interfere constructively. Our theoretical mechanism, illustrated in Fig.~\ref{fig:3}a, is consistent with the experimental observations (Fig.~\ref{fig:3}c). The remaining deviation between the calculated and measured major-axis orientations may arise from channel-dependent dephasing times, which were approximated in the calculations by a single value equal to one-quarter of an optical cycle.

To further test whether the chiral THG originates from multi-channel coherent superposition, we performed additional measurements at a driving wavelength of 4,000 $\mathrm{nm}$ using a separate experimental setup (Supplementary Section S1). In this case, the measured THG is nearly linearly polarized and remains almost parallel to the incident polarization for all orientation angles $\theta$ (Fig.~\ref{fig:4}e). In contrast to the 1,920 $\mathrm{nm}$ case, transitions between Weyl bands (group II) dominate the response, with intensities substantially larger than those of the other channels (Fig.~\ref{fig:4}b and c). This difference arises because the lower photon energy preferentially excites transitions near the Weyl points, whereas higher-lying bands become difficult to access. With fewer dominant channels, the conditions required to generate ellipticity through coherent superposition are no longer satisfied. The two channels in group II share similar polarization directions and phases (Supplementary Section S10) and therefore produce a linearly polarized harmonic response (Fig.~\ref{fig:4}d). Under these conditions, the process can be approximated as an effective two-band transition (Fig.~\ref{fig:4}a), which is much less likely to generate chiral harmonics.

\section{Discussion}
{Under ideal time-reversal dynamical symmetry and without another dynamical symmetry allowing elliptical polarization, a linearly polarized driver generates only linearly polarized harmonics in the single-channel Floquet description, as in the effective two-band case\cite{DynamicSymmInFloquet2017}. In real HHG, ionization, finite pulse duration, and dephasing can perturb or effectively invalidate this rule. Multichannel interband dynamics can, however, strongly relax the time-reversal-based polarization constraint and produce elliptical harmonics\cite{Linear2ellipticalHHGGaAs2023}. Thus, multichannel coherent superposition in a non-perturbative, time-reversal-symmetric system goes beyond the conventional Floquet dynamical-symmetry framework, yielding elliptically polarized harmonics with substantial ellipticity and well-defined helicity, and clarifying the roles of intra- and interband processes under time-reversal symmetry.}

{The central mechanism identified here is the coherent superposition of multiple interband transition channels with distinct transition-dipole orientations and emission phases under a single linearly polarized driver. This distinguishes the present effect from previous chiral HHG in GaAs under linear driving, where the ellipticity is attributed to interference between emission channels of different nonlinear orders and is weak for THG\cite{2022SekiguchiPrbGaAsLinear2Elliptical}. In NbP, by contrast, pronounced ellipticity already emerges at low harmonic orders in the deeply nonperturbative regime; the effect is therefore not a simple crossover phenomenon. The gapless multiband structure of NbP is an enabling factor because it provides strong optical nonlinearity, small direct gaps, and many optically accessible interband channels, including Weyl and non-Weyl bands. However, gaplessness alone is insufficient: near-circular THG requires two orthogonal harmonic-current components with comparable amplitudes and a relative phase close to $\pi/2$, which is not guaranteed by Weyl nodes. Our SBEs show that 16 of 28 interband channels contribute non-negligibly and that no small subset dominates; the essential requirement is multichannel coherent superposition.} 
{This is reinforced by gapless graphene and previous THG studies of NbP. In graphene, theory predicts only weak low-order ellipticity and traces its order dependence to the transition-dipole structure near the Dirac points, and pronounced elliptically polarized harmonics from pristine gapless graphene under a single linearly polarized driver have not been reported experimentally \cite{Dong2021TheoryLinear2EllipticalGraphene,Zurron2019TheoryLinear2EllipticalGraphene}. In NbP thin films, efficient THG has been linked to electronic states beyond the Weyl nodes, but this concerns THG efficiency rather than harmonic helicity \cite{tilmann2022NbPFilm3HG}. Thus, neither gapless dispersion alone nor a two-band Dirac structure is sufficient; multiple coherent interband channels with distinct polarizations and phases are essential for the NbP response calculated under the present driving conditions. Consistently, 4,000$\mathrm{nm}$ driving reduces the number of accessible interband channels, makes the response effectively two-band-like, and strongly suppresses the ellipticity.}

{Having established multichannel interband coherent superposition as the essential mechanism, we now clarify that topology, Berry curvature, and spin--momentum locking are not prerequisites for the observed harmonic helicity; at most, they can modify the relevant transition channels. In time-reversal-symmetric NbP, the Berry curvature satisfies 
$\Omega(-\mathbf{k}) = -\Omega(\mathbf{k})$, and within the conventional semiclassical intraband description the Berry-curvature-induced anomalous velocity produces even-order transverse harmonics under monochromatic linearly polarized excitation rather than directly generating the observed third-harmonic helicity; moreover, Weyl nodes of opposite chirality are required by the vanishing net topological charge, and their Brillouin-zone-integrated contributions may partially compensate. Spin--momentum locking and spin textures can modify optical transition strengths and selection rules, but they do not intrinsically impose the amplitude balance and relative phase required for circular polarization. The spin distributions and dipole similarities in Supplementary Sections S4 and S5 therefore serve to characterize and group the relevant transition channels, rather than to establish spin--momentum locking as an independent source of harmonic helicity. A possible additional topology-specific contribution is not separately quantified and remains an open question. We therefore expect the general mechanism---multichannel interband coherent superposition under a single linearly polarized driver---to apply beyond NbP to other gapless or small-gap multiband systems with strong optical nonlinearity and sufficiently rich interband transition channels, and not to be inherently restricted to topological Weyl semimetals.}

We can now summarize the key conditions for generating intense, highly elliptical HHG using only a single linearly polarized driving field. First, the crystal should exhibit strong optical nonlinearity to ensure efficient harmonic emission. Second, the material should possess a complex band structure and relatively small bandgaps near the Fermi level. Such a multiband electronic structure provides multiple interband transition channels that can coherently superpose with each other. 
{In NbP, the gapless multiband structure is an enabling rather than sufficient condition: it facilitates strong interband excitation and multiple channels, but near-circular THG additionally requires comparable orthogonal harmonic amplitudes and a relative phase close to $\pi/2$.}
{Third, the formation of chiral harmonics requires several emission pathways with comparable strengths. Phenomena that may accompany this condition include an abrupt rotation of the major axis of the harmonic polarization ellipse or a kink in the pump-intensity dependence\cite{2022SekiguchiPrbGaAsLinear2Elliptical}. Such features are consistent with competition among emission pathways. In our case, the channel-resolved SBEs provide the microscopic support, while the kink and the angular-pattern evolution serve as supporting phenomenology.}
Fourth, chiral harmonics are most likely to emerge when the driving-field orientation is away from both high-symmetry crystal directions and the orientations that maximize the harmonic yield. Although our study focuses on the Weyl semimetal NbP, in which the relevant multichannel processes are interband in nature, we expect these general guidelines to apply more broadly. The generality of the mechanism is supported by both experiment\cite{2022SekiguchiPrbGaAsLinear2Elliptical} and theory\cite{Zurron2019TheoryLinear2EllipticalGraphene}, which have likewise shown elliptical harmonic emission under linear excitation at low-symmetry orientations and highlighted the central role of coherent superposition. Establishing these design principles for chiral harmonic generation will aid the development of ultrafast light sources with tunable ellipticity.

Overall, by combining experimental measurements with SBEs, we demonstrate chiral harmonic generation in NbP driven by a single linearly polarized laser field and identify multichannel coherent superposition as the key microscopic mechanism. The chirality of the emitted harmonics also encodes important information about multiband transition pathways in a time-reversal-symmetric system. 
{The key factors identified in this work for chiral harmonic generation—strong nonlinearity, small direct gaps with multiple bands near the Fermi level, several interband channels with comparable amplitudes but distinct polarization and phase, and driving orientations away from high-symmetry directions and intensity maxima—can guide the construction of screening descriptors and the choice of calculation conditions in high-throughput Wannier–SBEs, possibly AI-accelerated, to rank candidate materials by computed ellipticity and channel-resolved phase or amplitude.}
More broadly, such advances will also facilitate the study of chirality-dependent light–matter interactions.

\section{Materials and Methods}

\subsection{Sample preparation}

The NbP single crystals were grown by the chemical vapor transport (CVT) method. A stoichiometric ratio of niobium powders and phosphorus pieces, together with iodine of 5 mg/ml as the transport agent, were sealed in an evacuated quartz ampoule. The ampoule was very slowly heated to avoid an explosion and finally exposed to a temperature gradient of 1000°C to 1025°C where the CVT proceeded for two weeks, then naturally cooled down to room temperature. NbP crystals with a size of $\sim$2 mm with regular shapes and shiny surfaces were obtained.

\subsection{HHG experimental setup}

The experiment was taken in a reflection spectroscopy geometry under ambient condition. Using 800$\mathrm{nm}$, 35fs, 1kHz Ti-sapphire femtosecond laser as source to pump the Spectra-Physics TOPAS Prime optical parametric amplifier (OPA), we produced 1,920$\mathrm{nm}$ idler beam. The NbP single crystal was oriented with \textit{b} axis along the horizontal direction. The pumping laser was focused onto the sample by N-BK7 lens (125-mm focal length), with a small incident angle ($\sim 8^{\circ}$) This angle was sufficiently small that the influence of oblique incidence was negligible (i.e., the THG contribution from the (010) surface could be ignored). Two 1500$\mathrm{nm}$ long-pass filters were used to ensure a pure 1,920$\mathrm{nm}$ incident light. 
We utilized the combination of quarter wave plate and polarizer instead of a single half wave plate to control the orientation angle $\theta$, where the fast axis of quarter wave plate was $45^{\circ}$ relative to the original 1,920$\mathrm{nm}$ linear light's polarization direction in order to generate circular light. That is because half wave plate may bring in retardation and cause ellipticity which we do not want, while polarizer with high extinction ratio ($\geq$800:1 for 1,920 $\mathrm{nm}$, Thorlabs WP25M-UB1) is able to ensure the linearity of our incident light.
Short pass filters (532$\mathrm{nm}$ short pass filter for measuring 4HG and 5HG; 750$\mathrm{nm}$ short pass filter for measuring THG) were used after the sample, as small signals might be overwhelmed by larger THG and 2HG signals. The HHG spectra were detected by a commercial spectrometer (wavelength 200-900$\mathrm{nm}$, ideaoptics).
The 4,000$\mathrm{nm}$ setup will be introduced in Supplementary Section S1.

\subsection{Calculation method}

The theoretical analysis and numerical simulations were performed by solving the SBEs in the Houston representation \cite{Wilhelm2021,Golde2008,Vampa2014,Li2019,Yue2020}. Atomic units are used throughout unless otherwise specified. The SBEs describe the laser-driven evolution of the single-particle density matrix $\rho_{nm}^{\mathbf{k}}(t)$ as:
\begin{align}
i\partial_t \rho_{nm}^{\mathbf{k}(t)}(t) &= \left[ \varepsilon_m^{\mathbf{k}(t)} - \varepsilon_n^{\mathbf{k}(t)} - \frac{i(1-\delta_{nm})}{T_2} \right] \rho_{nm}^{\mathbf{k}(t)}(t) \nonumber \\
&\quad - \mathbf{E}(t) \cdot \sum_l \left[ \mathbf{d}_{ln}^{\mathbf{k}(t)}\rho_{lm}^{\mathbf{k}(t)}(t) - \mathbf{d}_{ml}^{\mathbf{k}(t)}\rho_{nl}^{\mathbf{k}(t)}(t) \right],
\label{eq:SBEs}
\end{align}
Here, $\mathbf{k}(t) = \mathbf{K} + \mathbf{A}(t)$ denotes the laser-dressed crystal momentum, $\mathbf{K}$
is the canonical crystal momentum, $\mathbf{A}(t)$ is the vector potential of the
driving field, $\varepsilon_{n}^{\mathbf{k}}$ is the energy of the $n^{th}$ band at $\mathbf{k}$,
and $T_2$ is the phenomenological dephasing time. The dipole matrix element
$\mathbf{d}_{nm}^{\mathbf{k}} = i \langle u_{n,\mathbf{k}} | \nabla_\mathbf{k} | u_{m,\mathbf{k}} \rangle$
is evaluated from the cell-periodic Bloch functions
${u_{n,k}}$ and separates into diagonal ($n = m$) and off-diagonal ($n \neq m$) components,
associated with intraband motion and interband transitions, respectively.

First-principles calculations were carried out within density functional
theory using the Vienna Ab initio Simulation Package (VASP) \cite{VASP}.
Structural relaxation and self-consistent electronic-structure
calculations employed the Perdew--Burke--Ernzerhof (PBE)
exchange-correlation functional and projector-augmented-wave (PAW)
pseudopotentials, with SOC included throughout.
Brillouin-zone integrations used a uniform $10 \times 10 \times 14$ Monkhorst--Pack
$k$-point mesh, and the plane-wave kinetic-energy cutoff was set to $400$ eV.

Maximally localized Wannier functions (MLWFs) were generated using
Wannier90 \cite{Wannier90} and WannierTools \cite{WannierTools}. The resulting Wannier
Hamiltonian provided an efficient and gauge-consistent route to the
transition dipole matrix elements required for simulating HHG dynamics
\cite{Silva2019}. For the time-dependent calculations, the Brillouin zone was
sampled on a dense $600 \times 1000$ grid to resolve the band dispersion and
wavefunction. The MLWF construction yielded an orthogonal and complete
Bloch basis in a smooth Wannier gauge, thereby minimizing numerical
discontinuities between neighbouring $k$ points and enabling stable
evaluation of the dipole matrix and subsequent time propagation. The
static Bloch basis was determined by the initial Hamiltonian, whereas
the Houston basis followed the adiabatic evolution of these states under
the laser field.

The SBEs were integrated with a fourth-order Runge--Kutta scheme using a
time step of $0.15$ a.u., from which the time-dependent carrier
populations in the adiabatic Houston representation were obtained. The
total current was then resolved into interband and intraband
contributions as:
\begin{align}
\mathbf{J}_{\mathrm{inter}}(t) &= - \sum_{\mathbf{K} \in \mathrm{BZ}} \sum_{n \neq m} \rho_{nm}^{\mathbf{k}(t)}(t) \, \mathbf{p}_{nm}^{\mathbf{k}(t)}, \\
\mathbf{J}_{\mathrm{intra}}(t) &= - \sum_{\mathbf{K} \in \mathrm{BZ}} \sum_{n} \rho_{nn}^{\mathbf{k}(t)}(t) \, \mathbf{p}_{nn}^{\mathbf{k}(t)}.
\end{align}
Here, the momentum operator is given by
$\widehat{p}(\mathbf{k}) = \partial_{\mathbf{k}}\widehat{H}(\mathbf{k})$. In the Houston
representation, the corresponding matrix element is
$\mathbf{p}_{nm}^{\mathbf{k}(t)} = \langle u_{n}^{\mathbf{k}(t)} | \widehat{p} | u_{m}^{\mathbf{k}(t)} \rangle$
and is evaluated as:
\begin{align}
\mathbf{p}_{nm}^{\mathbf{k}(t)} &= i \left( \varepsilon_{n}^{\mathbf{k}(t)} - \varepsilon_{m}^{\mathbf{k}(t)} \right) \mathbf{d}_{nm}^{\mathbf{k}(t)}, \quad n \neq m, \\
\mathbf{p}_{nn}^{\mathbf{k}(t)} &= \nabla_{\mathbf{k}} \varepsilon_{n}^{\mathbf{k}(t)}.
\end{align}

{The calculation was performed over a total window of 184 fs, the driving field was modeled with a Gaussian envelope of FWHM 50 fs (approximately 8 optical cycles at 1,920 $\mathrm{nm}$). }

Finally, the HHG spectra were calculated by Fourier transforming the
total current. The simulated spectra were then compared quantitatively
with the experimental data, providing a stringent assessment of the
theoretical description.


\section*{Acknowledgments}
\subsection*{Funding}
  This work was supported by the National Natural Science Foundation of China (Grant No. 12488201, 12204520), the National Key Research and Development Program of China (Grant Nos. 2024YFA1408701, 2022YFA1403901, 2025YFA1411504, 2023YFA1607400), the Youth Innovation Promotion Association of CAS (No. 2021008).
\subsection*{Author Contributions} 
  R.X. Ding, L. Yue, T. Dong and N.L. Wang conceived the project. R.X. Ding performed the HHG measurements. C. Qian carried out the SBEs. Z.L. Li grew the NbP crystals. R.X. Ding, C. Qian, L. Yue, T. Dong, and N.L. Wang interpreted the data and wrote this paper, with inputs from all the authors.
\subsection*{Data availability}
  All data needed to evaluate the conclusions in the paper can be offered upon request.


\section*{Supplementary materials}
Supplementary Sections S1 to S17\\
Figs. S1 to S14\\
Table S1
\newpage


\renewcommand{\thefigure}{S\arabic{figure}}
\renewcommand{\thetable}{S\arabic{table}}
\renewcommand{\theequation}{S\arabic{equation}}
\renewcommand{\thepage}{S\arabic{page}}
\setcounter{figure}{0}
\setcounter{table}{0}
\setcounter{equation}{0}
\setcounter{page}{1} 


\begin{center}
\section*{Supplementary Materials for\\ Chiral high harmonic generation via coherent superposition of multi-channel interband transitions}

Ruoxuan~Ding \textit{et al.}
\\ 
\small Corresponding authors: Chen Qian, cqian@iphy.ac.cn; Zhilin Li, lizhilin@iphy.ac.cn; Li Yue, yueli@baqis.ac.cn; Nan-Lin Wang, nlwang@sjtu.edu.cn\\
\end{center}

\subsubsection*{This PDF file includes:}
Materials and Methods\\
Supplementary Sections S1 to S17\\
Figs. S1 to S14\\
Table S1\\

\newpage


\subsection*{Supplementary Section S1. Optical setup of 4 \textmu m experiment}

Similar to the setup with fundamental wavelength at 1,920 $\mathrm{nm}$, 4,000 $\mathrm{nm}$ HHG experiment also adopts a reflective setup, shown in Supplementary Fig.~\ref{fig:S1}.
The signal and idler beam from OPA are used for differential frequency generation in $\mathrm{AgGaS_2}$ crystal to generate 4 \textmu m light.
Signal beam is tuned to an appropriate wavelength(approximately 1,333$\mathrm{nm}$) satisfying conservation of photon energy as follows: 
\begin{equation}
	\frac{1}{800\mathrm{nm}} = \frac{1}{\lambda_{\mathrm{Signal}}} + \frac{1}{\lambda_{\mathrm{Idler}}}
\end{equation}
\begin{equation}
	\frac{1}{\lambda_{\mathrm{DFG}}} = |\frac{1}{\lambda_{\mathrm{Signal}}} - \frac{1}{\lambda_{\mathrm{Idler}}}|
\end{equation}
A Ge plate (as a long pass filter) and  a bandpass filter with central wavelength at 4 \textmu m are used to filter OPA light and THG generated in optics, purifying the 4 \textmu m driver light. 
A combination of $45^{\circ}$ quarter waveplate plus polarizer is also adopted instead of a half waveplate, in order to make sure the linearity of incident beam.
A $\mathrm{CaF_2}$ lens is used instead of a N-BK7 lens before the sample, for higher transmission rate of mid-infrared light. But the lens after the sample is still a N-BK7 lens, in order to filter 4 \textmu m fundamental light, suppressing THG generated in fiber. The definition of $\theta$ and $\phi$ are the same as those in the main text.

{In both setups of 1,920$\mathrm{nm}$ and 4,000$\mathrm{nm}$, the polarizer and wave plate angles were calibrated in situ using a horizontal linearly polarized laser. The zero of $\theta$ was defined by the \textit{b} axis of the same NbP crystal, determined by Laue diffraction. Residual mounting offsets were checked and minimized using the expected mirror symmetry of (101) face. Thus the angular scales of the 1,920$\mathrm{nm}$ and 4,000$\mathrm{nm}$ data sets are consistent.}

\begin{figure}[htbp]
    \begin{center}
	   \includegraphics[page=1, width=1\textwidth, trim=2in 2in 2in 2in, clip]{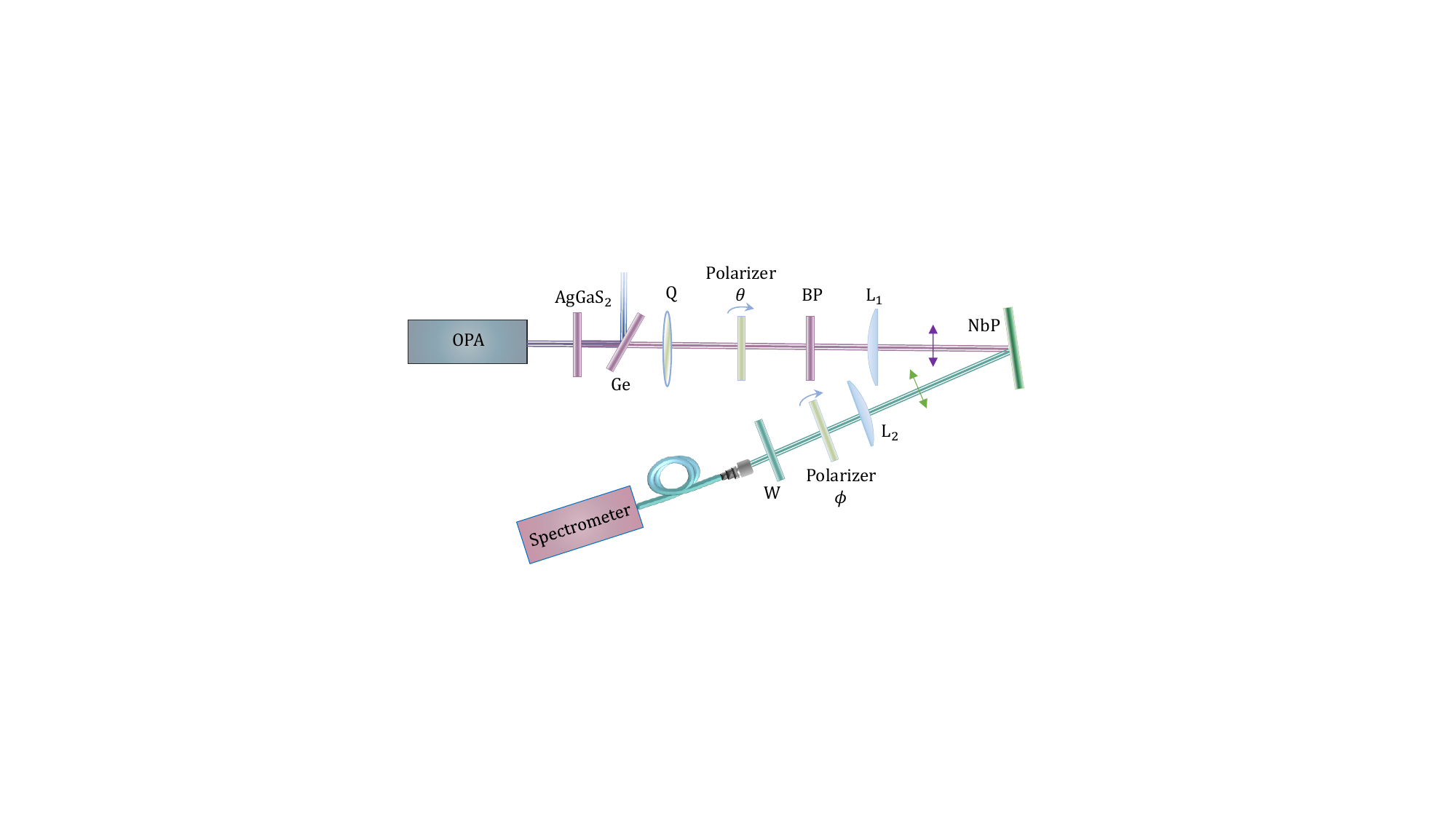}
	   \caption{\textbf{Optical setup of 4 \textmu m experiment.} Q: quarter wave plate, with fast axis along $45^{\circ}$; BP: bandpass filter with a center wavelength of 4 \textmu m; $\mathrm{L_1}$: $\mathrm{CaF_2}$ lens, f = 100mm; $\mathrm{L_2}$: N-BK7 lens, f = 50mm; W: N-BK7 window.}
	   \label{fig:S1}
    \end{center}
\end{figure}

\subsection*{Supplementary Section S2. The first Brillouin zone of NbP}
 
In Supplementary Fig.~\ref{fig:S2}, the first Brillouin zone (BZ) in momentum space for the primitive unit cell is illustrated. 
{The $C_4$ axis along $k_z$ passes through $\Gamma$ and Z, inherited from the fourfold screw axis of the tetragonal I4$_1$md structure. }
The pink and green planes represent mirror planes perpendicular to the ${k_y}$ and ${k_x}$ direction respectively; the capital letters indicate high-symmetry points in the Brillouin zone, and the purple arrows indicate the band path. {a*, b*, and c* denote the primitive reciprocal lattice vectors corresponding to the real-space lattice vectors a, b, and c.}

\begin{figure}[htbp]
    \begin{center}
	   \includegraphics[page=2, width=1\textwidth, trim=2.3in 2.5in 2.8in 2.2in, clip]{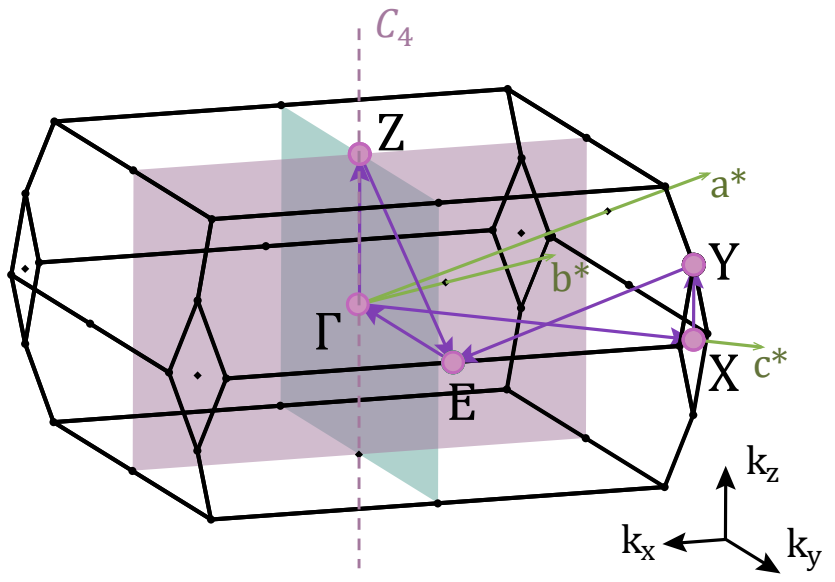}
	   \caption{\textbf{The first Brillouin zone of NbP.} The $C_4$ axis is along $k_z$, passing through $\Gamma$ and Z. Pink and green planes are mirror planes perpendicular to $k_y$ and $k_x$, respectively. Capital letters denote high-symmetry points, purple arrows denote the band path, and a*, b*, c* are the primitive reciprocal-lattice vectors.}
	   \label{fig:S2}
    \end{center}
\end{figure}

\subsection*{Supplementary Section S3. Theoretical analysis: polarization constraint and multiband coherence in THG}
Here we establish the symmetry constraint for a time-reversal-symmetric crystal driven by a linearly polarized field. 
Within the two-band semiconductor Bloch framework, the photocurrent can be decomposed into intraband and interband contributions. 
The phase of the photocurrent contains the dynamical phase, the shift-vector-induced phase, and the transition-dipole phase difference. 
The interband and intraband currents share the same phase structure, so their coherent emission is governed by the same symmetry constraints \cite{qc2024HhgSymm}.

For a linearly polarized driving field,
\begin{equation}
    \mathbf{E}(t) = E_b(t) \hat{\mathbf e}_b ,
\end{equation}
we choose the temporal centre $t_c$ of the pulse such that
\begin{equation}
    E_b(t_c+\tau) = E_b(t_c-\tau).
\end{equation}
The vector potential can then be chosen as
\begin{equation}
    A_b(t_c+\tau) = - A_b(t_c-\tau).
\end{equation}
For the field-dressed crystal momentum
\begin{equation}
    \mathbf{k}(t) = \mathbf K + \mathbf A(t),
\end{equation}
time reversal maps the strong-field trajectory as
\begin{equation}
    \mathbf{k}(t_c-\tau) = -\mathbf{k}(t_c+\tau).
\end{equation}

In a below-gap response regime, real carrier injection is weak. 
The population difference can therefore be treated as approximately covariant under the above dynamical time-reversal operation,
\begin{equation}
    f_{cv}(-\mathbf K, t_c-\tau) \simeq f_{cv}(\mathbf K, t_c+\tau).
\end{equation}
Under the combined operation of time reversal and a half-cycle time translation,
\begin{equation}
    \hat U = \hat T \cdot \hat{\tau}_2 ,
\end{equation}
the interband current transforms as \cite{qc2024HhgSymm}
\begin{equation}
\begin{split}
    \hat U J_{nm}^a(\mathbf K,t) \hat U^\dagger
    &= J_{nm}^a(-\mathbf K, \hat{\tau}_2 t) \\
    &= - J_{nm}^a(\mathbf K,t) 
    \exp \Big[ 2i \big( S_{\rm shift}(\mathbf K,t,t') + S_{\rm TDP}[\mathbf k(t)] \big) \Big].
\end{split}
\end{equation}

This relation shows that time-reversal-related channels are connected by a well-defined phase relation rather than by a purely scalar sign change. 
In the below-gap limit, after summing over the Brillouin zone, the $\mathbf K$ and $-\mathbf K$ channels form approximately conjugate pairs. 
As a result, each Cartesian component of the emitted current satisfies
\begin{equation}
    J^a(t_c-\tau) \simeq - J^a(t_c+\tau).
\end{equation}

The $N$-th harmonic amplitude is defined as
\begin{equation}
    J_N^a = \int dt\, W(t-t_c) J^a(t) e^{iN\omega_0 t},
\end{equation}
where $W(t-t_c)$ is a real even temporal window. 
Since $J^a(t_c+\tau)$ is approximately odd in $\tau$, the cosine component of the Fourier integral vanishes, and one obtains
\begin{equation}
    J_N^a = e^{i\Phi_N} C_N^a, \qquad C_N^a \in \mathbb{R}.
\end{equation}
Here $\Phi_N$ is a common phase shared by all Cartesian components of the same harmonic order. Therefore, for the two components parallel and perpendicular to the driving field,
\begin{equation}
    J_N^\parallel = e^{i\Phi_N} C_N^\parallel, \qquad
    J_N^\perp = e^{i\Phi_N} C_N^\perp,
\end{equation}
with
\begin{equation}
    C_N^\parallel, C_N^\perp \in \mathbb{R}.
\end{equation}
Consequently,
\begin{equation}
    \operatorname{Im} \left[ J_N^\parallel J_N^{\perp *} \right] = 0,
\end{equation}
and the circular Stokes component vanishes,
\begin{equation}
    S_3^{(N)} = \frac{2 \operatorname{Im} \left[ J_N^\parallel J_N^{\perp *} \right]}{|J_N^\parallel|^2 + |J_N^\perp|^2} = 0.
\end{equation}

Thus, below-gap low-order harmonics emitted from a time-reversal-symmetric two-band system remain linearly polarized. Importantly, this does not require
\begin{equation}
    C_N^\perp = 0.
\end{equation}
Crystal anisotropy and the direction of transition dipoles can change the relative amplitudes of the two orthogonal components. The harmonic polarization axis may therefore rotate with respect to the incident linear polarization, with
\begin{equation}
    \tan \theta_N = \frac{C_N^\perp}{C_N^\parallel}.
\end{equation}

In gapless systems such as the Weyl semimetals, even low-order THG involves strong interband excitations, and the strict dynamic time-reversal symmetry of the populations is not maintained. In a two-band approximation, this produces a finite phase variation, resulting in a weak ellipticity. However, the THG signal does not reach circular polarization in this approximation.

When multiple interband channels are included, the third-harmonic current becomes a coherent sum over different pathways:
\begin{equation}
    J_3^a = \sum_\mu J_{3,\mu}^a,
\end{equation}
with orthogonal components
\begin{equation}
    J_3^x = \sum_\mu J_{3,\mu}^x, \quad J_3^y = \sum_\mu J_{3,\mu}^y.
\end{equation}

The circular component of THG is determined by the coherent cross term
\begin{equation}
    \operatorname{Im}[J_3^x J_3^{y*}] = \operatorname{Im} \Big[ \sum_{\mu,\nu} J_{3,\mu}^x J_{3,\nu}^{y*} \Big].
\end{equation}

Expressing the THG polarization in terms of complex amplitudes:
\begin{equation}
    J_3^x = |J_3^x| e^{i\phi_x}, \quad J_3^y = |J_3^y| e^{i\phi_y},
\end{equation}
the circular Stokes parameter reads
\begin{equation}
    S_3^{(3)} = \frac{2 |J_3^x| |J_3^y| \sin(\phi_x - \phi_y)}{|J_3^x|^2 + |J_3^y|^2}.
\end{equation}

If $\phi_x - \phi_y = 0$ or $\pi$, the THG remains linearly polarized. 
A nonzero phase difference produces a finite ellipticity, and only when 
$|J_3^x| = |J_3^y|$ and $\phi_x - \phi_y = \pm \pi/2$ does the THG become fully circularly polarized. This shows that the circular polarization of THG is a consequence of
multiband coherent interference between orthogonal components. From our calculation, two-band THG produces merely a small ellipticity. However, inclusion of multiband coherence generates pronounced near-circular THG emission. Therefore, multiband coherent interference is the essential mechanism behind circular polarization in linearly driven THG from Weyl semimetals.

\subsection*{Supplementary Section S4. Spin distribution in k-space}

We plot the spin distribution of the 8 bands involved in our SBEs in Supplementary Fig.~\ref{fig:S3}. In the left panel, the spin distribution of band V1, V3, C2 and C4 shows high similarity, as do band V2, V4, C1 and C3 in the right panel. This similarity favors higher transition probability, i.e., transitions between V1, V3, C2 and C4 (V2, V4, C1 and C3) are more likely to happen.

\begin{figure}[htbp]
    \begin{center}
	   \includegraphics[page=3, width=1\textwidth, trim=2in 2in 2in 2in, clip]{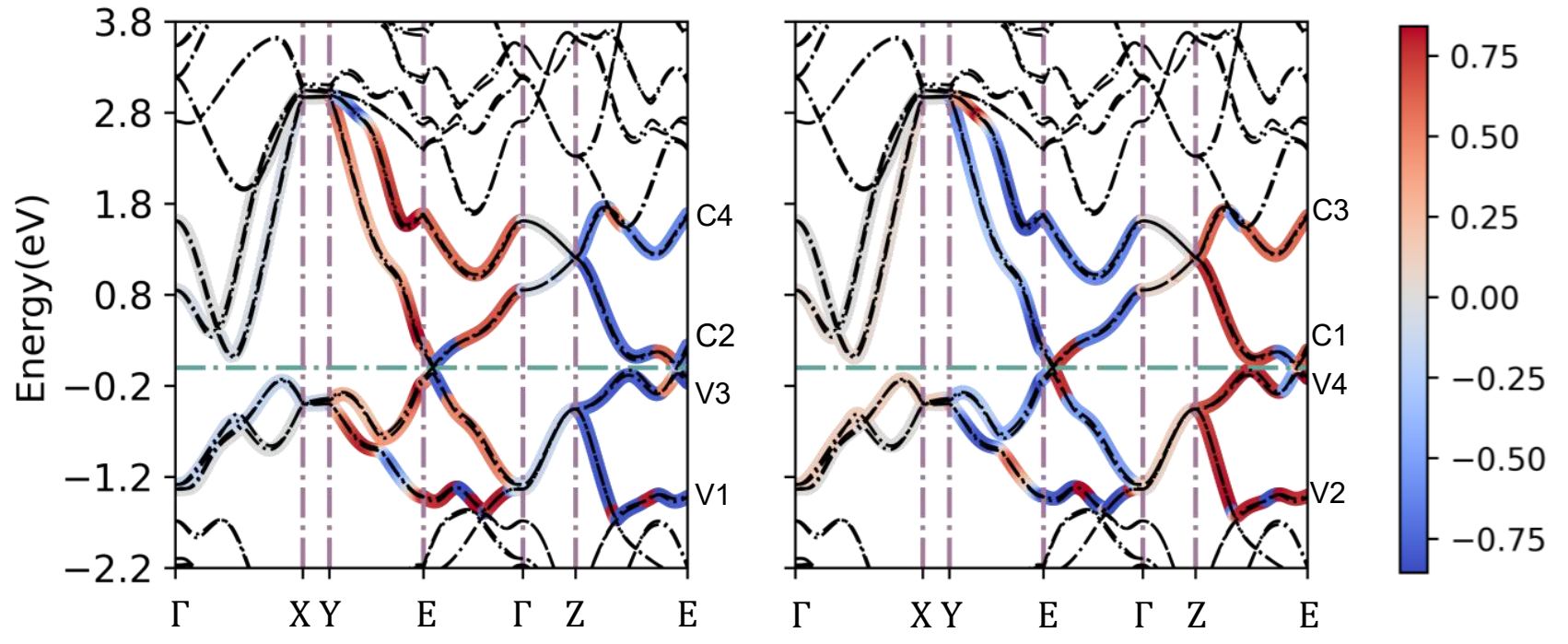}
	   \caption{\textbf{Spin distribution in k-space.} the spin distribution of band structure alone high symmetry paths. The shade color indicates the spin value, ranging from -1(blue) to +1(red). Left panel: spin distribution of band V1, V3, C2 and C4; right panel: spin distribution of band V2, V4, C1 and C3.}
	   \label{fig:S3}
    \end{center}
\end{figure}

\subsection*{Supplementary Section S5. Transition dipole distribution in k-space}

In Supplementary Fig.~\ref{fig:S4}, C3-V4 and C4-V3 (C1-V4 and C2-V3) show similar dipole distribution, as C3 and C4 (V3 and V4, C1 and C2) are subbands split by SOC. This similarity likely explains why these paired interband channels have almost the same polarization direction. Moreover, their dipole magnitude is greater than that of C3-V3 and C4-V4 (C1-V3 and C2-V4), that is because the valence bands and conduction bands in these four transition channels show similar spin distribution,  respectively, shown in Supplementary Fig.~\ref{fig:S3}, favoring higher transition probability.

\begin{figure}[htbp]
    \begin{center}
	   \includegraphics[page=4, width=1\textwidth, trim=2.3in 0.5in 2in 0.5in, clip]{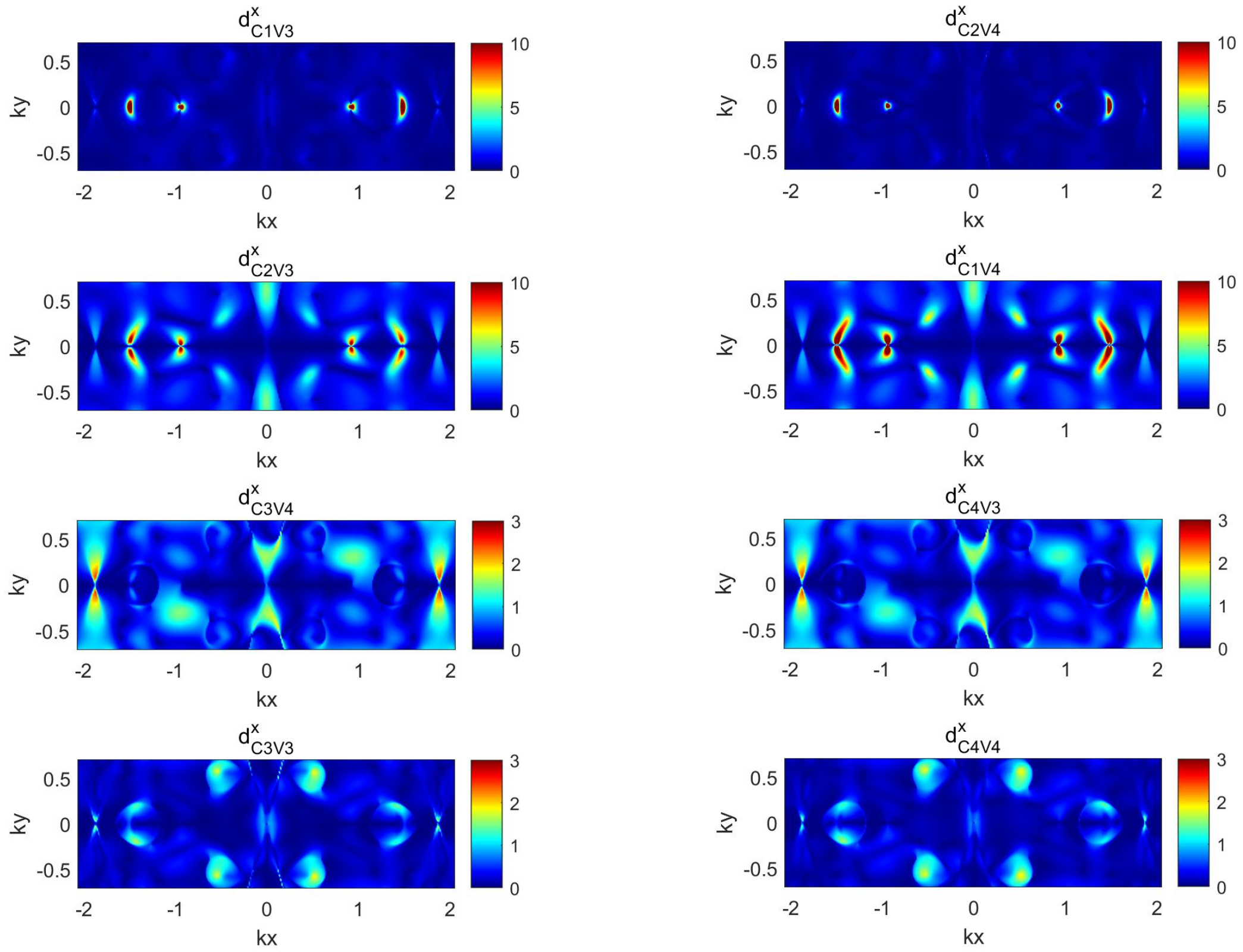}
	   \caption{\textbf{Transition dipole distribution in k-space.} The component of transition dipole along \textit{b} axis in real space ($d^x$) distribution in k-space of (101) face. $k_x$ indicates the reciprocal-space wave vector component corresponding to the real-space \textit{b} axis, $k_y$ indicates the reciprocal-space wave vector component corresponding to the axis perpendicular to \textit{b} axis in (101) face.}
	   \label{fig:S4}
    \end{center}
\end{figure}

\subsection*{Supplementary Section S6. Penetration depth, spot diameter and peak power density}

At 1,920 $\mathrm{nm}$, penetration depth (electric field decays to 1/e of its initial value) of electric field $z_E$ is 82$\mathrm{nm}$. This is calculated by utilizing Kramers–Kronig relations to get extinction coefficient $k$ from reflectivity measure in literature\cite{PhysRevB.98.195203}, then the penetration depth of intensity is $z_I = \frac{z_E}{2} =  \frac{1}{\alpha}$, where the absorption coefficient $\alpha = \frac{4\pi k}{\lambda}$.

We control the same peak power density for 1,920 $\mathrm{nm}$ and 4,000 $\mathrm{nm}$. For 1,920 $\mathrm{nm}$, the spot diameter is 184 \textmu m, peak power density $2.7\times10^{11}\,\mathrm{W\,cm^{-2}}$, with single pulse energy 2$\mu$J, pulse width 60fs. For 4,000 $\mathrm{nm}$, the spot diameter is 367.8 \textmu m, peak power density $2.5 \times10^{11}\,\mathrm{W\,cm^{-2}}$, with single pulse energy 13$\mu$J, pulse width 100fs.

\subsection*{Supplementary Section S7. Chirality of THG and 4HG}

To check the chirality of elliptical THG, we perform an additional polarization measurement using a quarter-wave plate (QWP). 
The QWP is put behind the sample, with its fast axis along the minor axis of THG ellipse ($127^{\circ}$) and the major axis of 4HG ellipse($1^{\circ}$). The polarization of THG and 4HG before and after the QWP is shown in Supplementary Fig.~\ref{fig:S5}a and b.
The ellipse of THG and 4HG can be converted to a near linear light, whose major axis is around $5^{\circ}$ and $153^{\circ}$.
The chirality can be determined by the angle $\alpha$, defined as the angle from the output linear polarization direction to the fast axis of the QWP, measured counterclockwise ($0^{\circ}<\alpha<180^{\circ}$).
Specifically, if $0^{\circ}<\alpha<90^{\circ}$, the light is left‑handed polarized; if $90^{\circ}<\alpha<180^{\circ}$, the light is right‑handed polarized. This rule applies regardless of whether the QWP fast axis is aligned with the major axis or the minor axis of the ellipse.
Here, the THG in Supplementary Fig.~\ref{fig:S5}a, with $\alpha = 122^{\circ}$, is right-handed; the 4HG in Supplementary Fig.~\ref{fig:S5}b, with $\alpha = 28^{\circ}$, is left-handed.

\begin{figure}[htbp]
    \begin{center}
	   \includegraphics[page=5, width=1\textwidth, trim=2in 1in 2.5in 1in, clip]{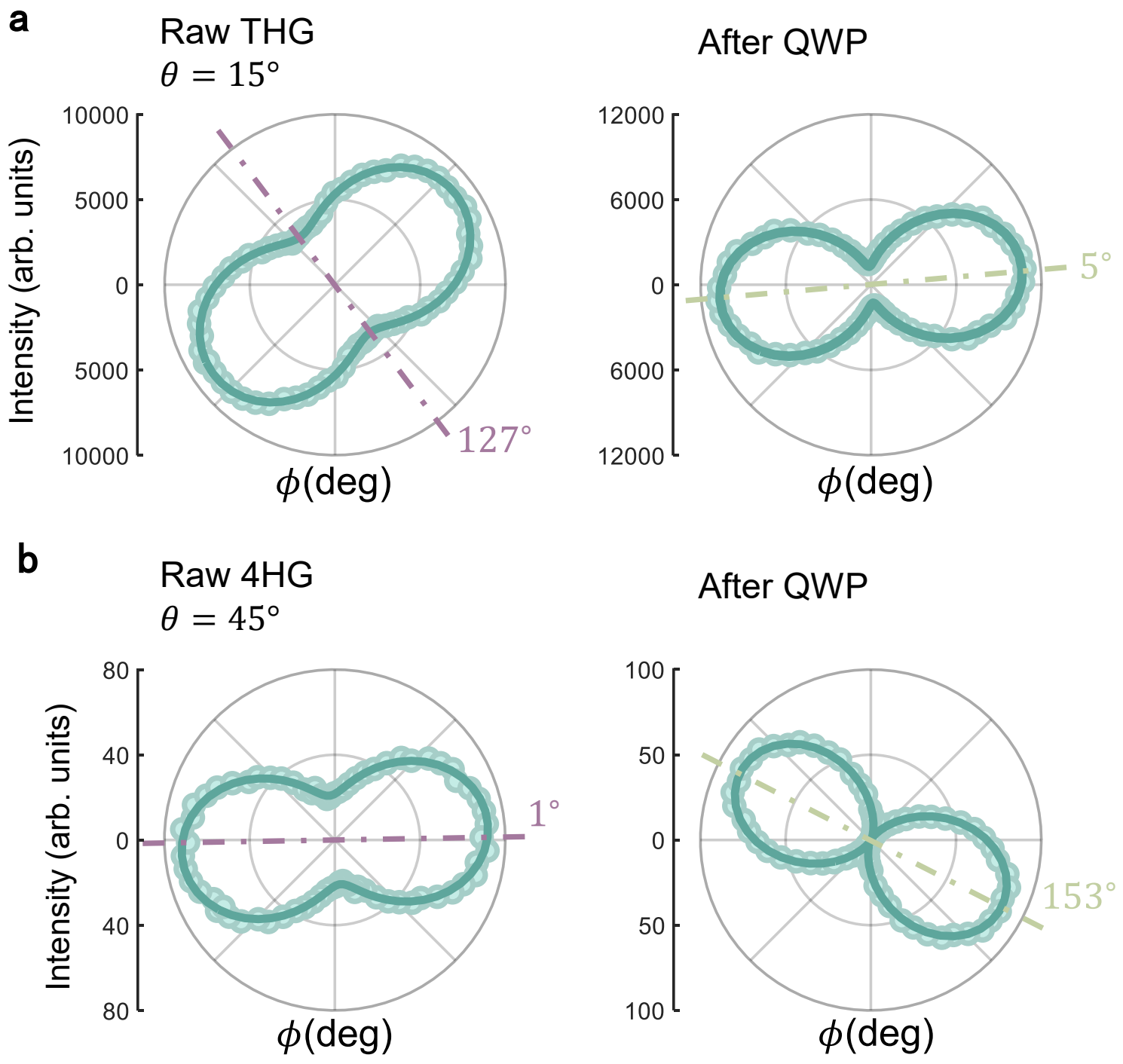}
	   \caption{\textbf{Chirality measurement.} We put a QWP behind the sample, to convert the chiral harmonic into a linear light. The fundamental light is $\sim$1,920 $\mathrm{nm}$, with peak power density $\sim2.6\times10^{11}\,\mathrm{W\,cm^{-2}}$. The green circles are experimental data, while the green solid curves are fitting lines.  The pink dashed line indicates the fast axis of QWP, and the yellow dashed line indicates the major axis of linear light. By examining the relative angle between them, we can determine that the THG and 4HG in \textbf{(a)} and \textbf{(b)} are right-handed and left-handed, respectively.}
	   \label{fig:S5}
    \end{center}
\end{figure}

\subsection*{Supplementary Section S8. Kinks in HHG yield's dependence on pump intensity}

\begin{figure}[htbp]
    \begin{center}
	   \includegraphics[page=6, width=1\textwidth, trim=2in 2in 2.5in 2.2in, clip]{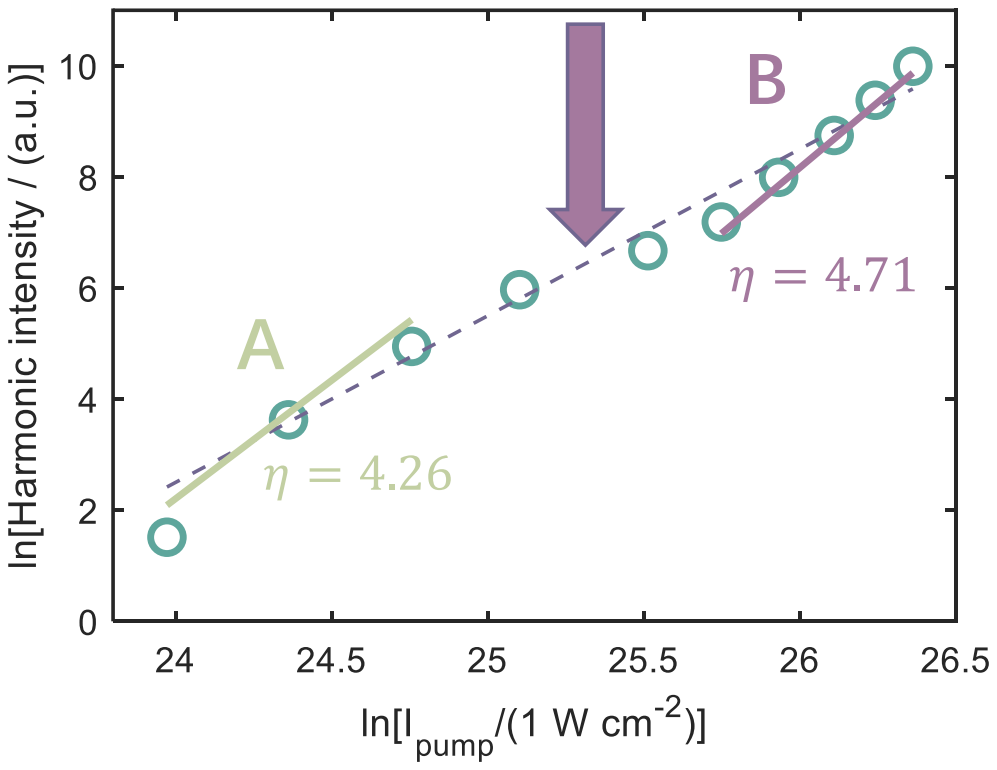}
	   \caption{\textbf{Logarithmic plot of THG yield with respect to pumping intensity.} A kink appear in pumping intensity dependent THG yield, marked by a purple arrow. We hypothesis that the region before and after the kink are dominated by A and B processes respectively, and fit them with the power-law equation $I \propto I^n_{\mathrm{pump}}$ to get the values of the n-factor, marked as $\eta$ with yellow and pink color in the figure. The dashed line indicates the $I^3_{\mathrm{pump}}$ scaling reflecting perturbative THG.}
	   \label{fig:S6}
    \end{center}
\end{figure}

Pump intensity dependent HH yield is usually used to tell whether the process is perturbative or non-perturbative: $n^{th}$ order harmonic deviating from perturbative power law scaling $I^{n}$ implies the process enter non-perturbative region. However, in some cases there may appear a unique feature in pump intensity dependent HH yield: a kink. It is often thought to be related to interference, two different emission process dominate separately the process before and after the kink region(at low and high pump intensity), but share similar strength at some pump intensity thus allow interference   to happen and cause a kink\cite{PrbWP2Calculation}\cite{2022SekiguchiPrbGaAsLinear2Elliptical}\cite{Kim2025Wp2Kinks}.
In our work, we also observe a kink in pumping intensity dependent THG yield in (1,0,1) face,(Supplementary Fig.~\ref{fig:S6}) which coincides with orientation dependent petal shape transition region. A and B process respectively dominate the regions before and after the kink, and show different HH anisotropy.

\newpage
\subsection*{Supplementary Section S9. Origin of even harmonic generation}
It is usually assumed that only odd harmonics can appear in the parallel direction in typical semiclassical model\cite{lv2021WeylWP2HHG}. However, parallel 2HG is also observed, which is not possible to be generated by intraband Berry curvature and usually attributed to interband transition in literature\cite{PrbWP2Calculation, 2024PRAhBNparaSHG}.
In our work, 4HG exhibit significant component along the direction of pumping laser (also exhibit high ellipticity, shown in Supplementary Fig.~\ref{fig:S7}), indicating indispensable role of interband transition.

\begin{figure}[htbp]
    \begin{center}
	   \includegraphics[page=7, width=1\textwidth, trim=2in 2.5in 3in 2in, clip]{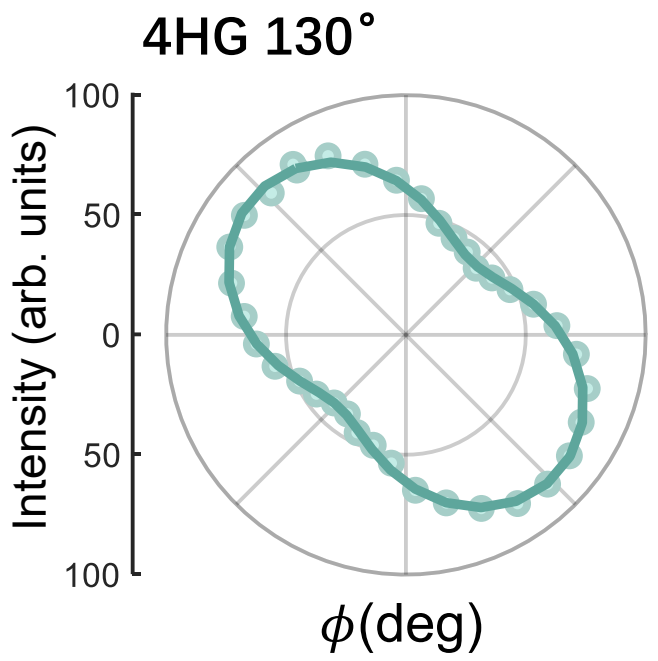}
	   \caption{\textbf{4HG of NbP at 1,920 $\mathrm{nm}$.} Polarization of 4HG at $\theta = 130^{\circ}$. The peak power density of the driving field is $2.7\times 10^{11}\,\mathrm{W\,cm^{-2}}$. }
	   \label{fig:S7}
    \end{center}
\end{figure}

\subsection*{Supplementary Section S10. Detailed analysis of interband transition channels at 1,920 $\mathrm{nm}$}

At 1,920 $\mathrm{nm}$, the top four interband transition channels by intensity are C4-V3, C2-V3, C3-V4 and C1-V4. It has already been explained physically in Supplementary Section S4, why these channels have high intensity. These four channels can be further categorized into group \uppercase\expandafter{\romannumeral1} (C3V4 and C4V3) and group \uppercase\expandafter{\romannumeral2} (C2V3 and C1V4) based on their electronic origins: transition between a Weyl band and a trivial conduction band and transition between Weyl bands. Within each group, the two pathways produce harmonics with similar phase and polarization direction, thus combine to produce harmonics with linear polarization. 
Actually, the introduction of SOC transforms the originally four-band system into eight bands. Although the number of transition channels increases, channels that are degenerate in the absence of SOC often exhibit high similarity in polarization and intensity when the SOC is weak. This is because the SOC‑induced mixing between spin up and down is small, so the transition dipole elements approximately retain the character of those from the spin‑conserved picture (i.e., up-spin electrons transit to up-spin holes, down-spin electrons transit to down-spin holes). The interband transition channels with broken spin degeneracy can be approximated as one transition channel (here we put them into the same group), further evidenced by similar transition dipole distributions (Supplementary Section S5). Channels from different groups, however, have distinct electronic origins and are likely to exhibit different phases and polarization, which favors generation of elliptical harmonic generation.
The polarization of group \uppercase\expandafter{\romannumeral1}, \uppercase\expandafter{\romannumeral2} and their addition at $\theta = 20^{\circ}$ is shown in Supplementary Fig.~\ref{fig:S8}. Although the polarizations of the two groups are perpendicular to each other, the phase difference between them is not close to $\pi/2$, not meeting the condition of generating harmonic with high ellipticity.
For a vertical two-band transition excited by a linear laser, the emitted HHG is expected to be linearly polarized. However, we note that harmonic generation calculated from one single channel can also show slight ellipticity. That could be due to excitations at distinct k-points of the same valence band, introducing slight phase difference. Nonetheless, this mechanism does not yield significant ellipticity, which still requires coherent superposition between multiple interband transition channels.

\begin{figure}[htbp]
    \begin{center}
	   \includegraphics[page=8, width=1\textwidth, trim=2in 2in 3in 2in, clip]{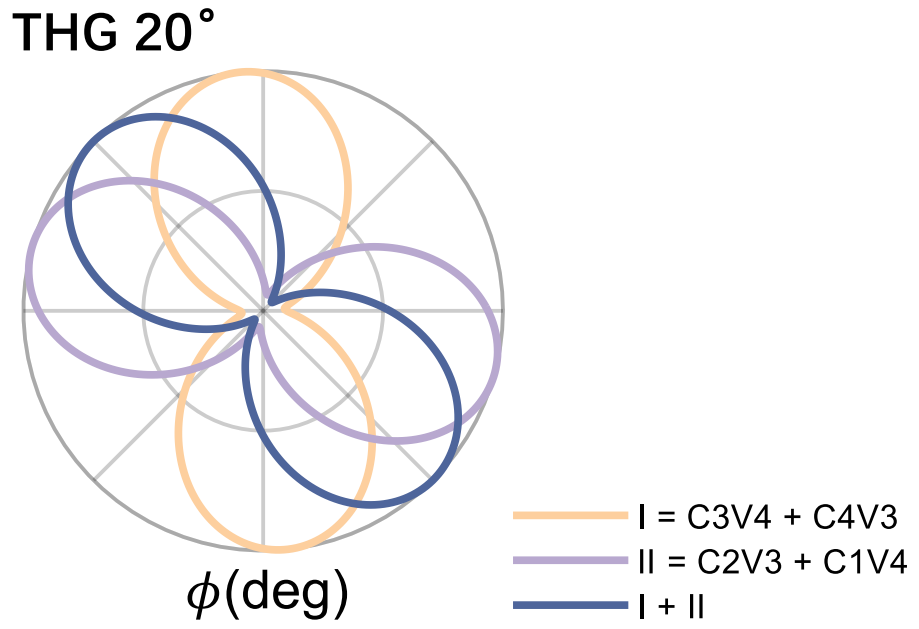}
	   \caption{\textbf{Polarization of four channels' addition.} Polarization of the top four channels by intensity(group \uppercase\expandafter{\romannumeral1} and \uppercase\expandafter{\romannumeral2}) and their coherent superposition (denoted as \uppercase\expandafter{\romannumeral1} + \uppercase\expandafter{\romannumeral2}) at $\theta = 20^{\circ}$, 1,920 $\mathrm{nm}$. }
	   \label{fig:S8}
    \end{center}
\end{figure}



\subsection*{Supplementary Section S11. Orientation dependence and ellipticity of 4HG and 5HG in NbP. }
\begin{figure}[h]
    \begin{center}
	   \includegraphics[page=15, width=1\textwidth, trim=1in 1.2in 1in 1in, clip]{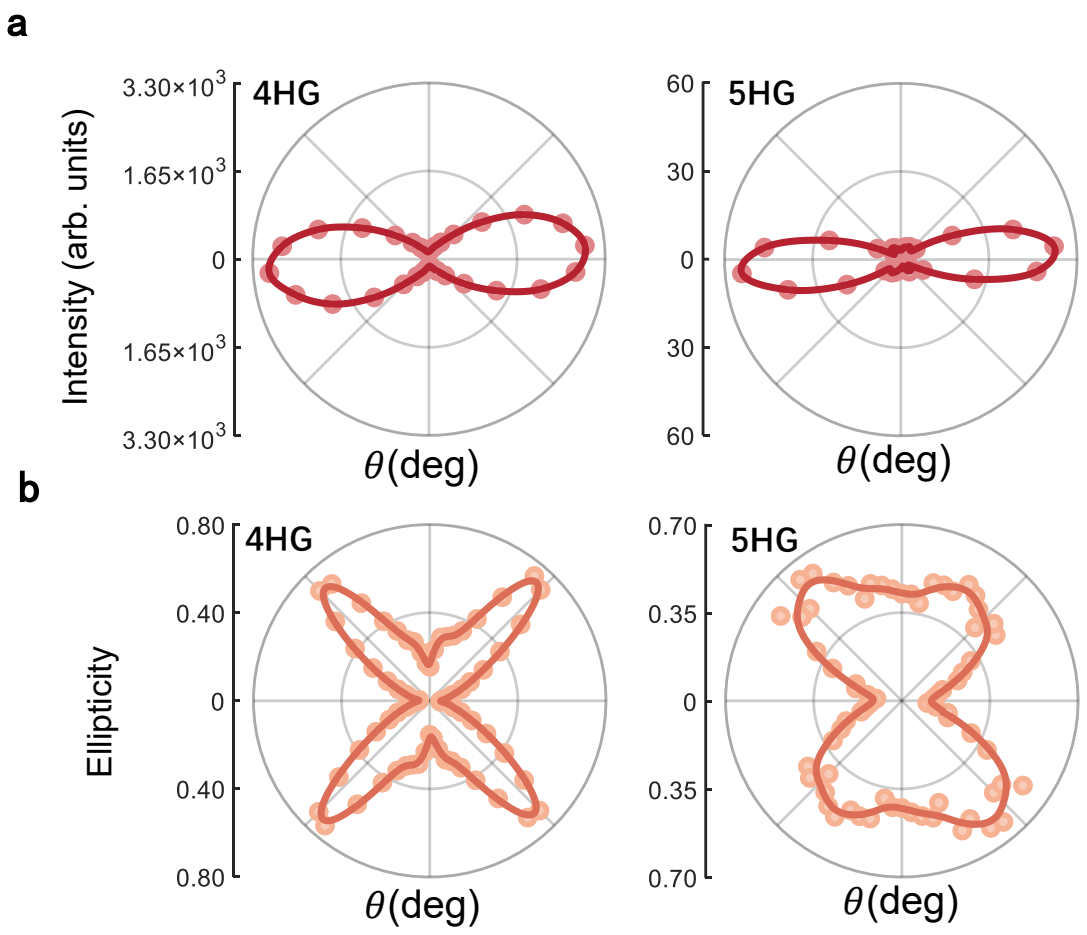}
	   \caption{\textbf{Orientation dependence and ellipticity of 4HG and 5HG in NbP.}\textbf{(a)} Orientation-dependent intensities of the fourth and fifth harmonics from the NbP (101) surface.
Experimental data are shown as red circles. Solid lines are generated by interpolation solely as a visual guide for the eye. \textbf{(b)}, Experimentally measured $\epsilon$ as a function of $\theta$. Solid lines are generated by interpolation solely as a visual guide for the eye. The measurements were
performed at a peak intensity of $\sim 2.7 \times 10^{11}\,\mathrm{W\,cm^{-2}}$.}
	   \label{fig:S15}
    \end{center}
\end{figure}

{Fig.~\ref{fig:S15} summarizes the orientation-dependent polarimetric response and ellipticity of 4HG and 5HG at a peak intensity of $\sim 2.7 \times 10^{11}\,\mathrm{W\,cm^{-2}}$ from the NbP (101) surface.
Figure~\ref{fig:S15}a shows the harmonic intensity as a function of $\theta$, with the polarizer P2 for HHG output removed. 
Fig.~\ref{fig:S15}b summarizes the ellipticity extracted from systematic polarization-resolved measurements as a function of $\theta$. }

\newpage
\subsection*{Supplementary Section S12. Polarization-state decomposition of intraband and interband contributions to THG}

\begin{figure}[htbp]
    \begin{center}
	   \includegraphics[page=10, width=1\textwidth, trim=1in 1.7in 1in 1.5in, clip]{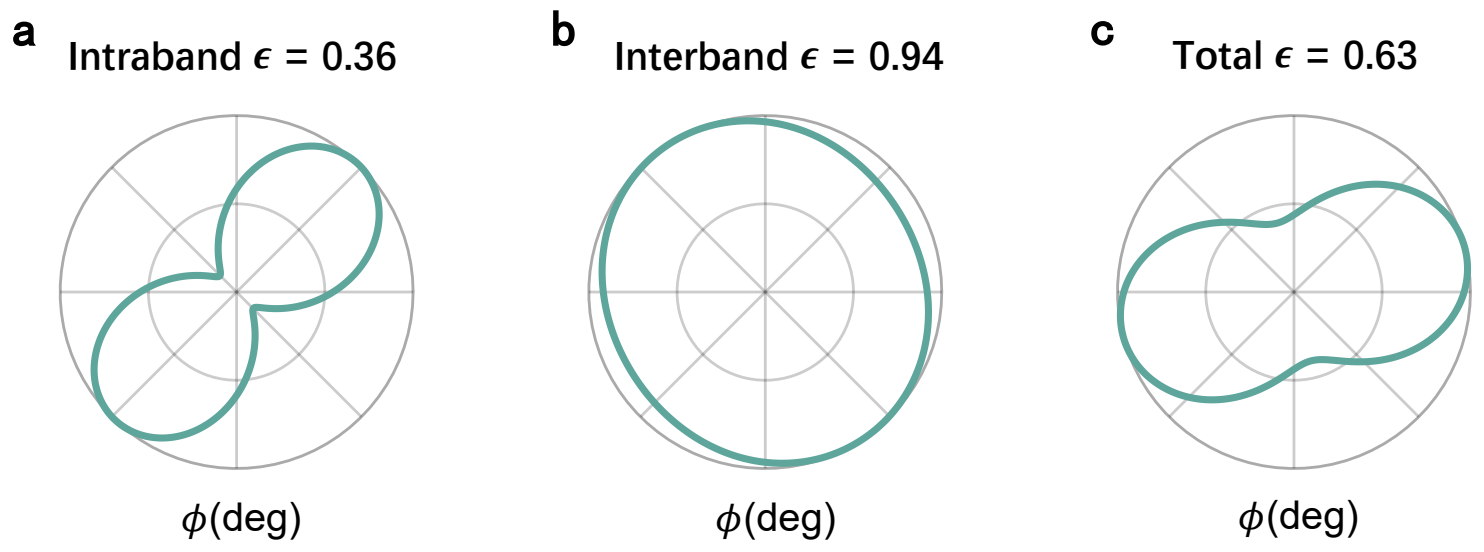}
	   \caption{\textbf{Calculated THG polarization states obtained from intraband-only, interband-only, and total-current contributions.} \textbf{(a)}, Calculated polarization ellipse of THG at $\theta = 15^{\circ}$ obtained from intraband-only. \textbf{(b)}, Calculated polarization ellipse of THG at $\theta = 15^{\circ}$ obtained from interband-only. \textbf{(c)}, Calculated polarization ellipse of THG at $\theta = 15^{\circ}$ obtained from total current.}
	   \label{fig:S10}
    \end{center}
\end{figure}

{To verify that the chiral THG originates from interband transitions rather than intraband acceleration or interference between interband and intraband contribution, we decomposed the calculated THG current into intraband-only and interband-only at a representative driving condition ( (101) face, $\lambda = 1,920\mathrm{nm}$, $\theta = 15^{\circ}$, peak intensity $\sim 2.7\times10^{11}\,\mathrm{W\,cm^{-2}}$). 
Fig.~\ref{fig:S10} compares the THG polarization ellipses obtained from the intraband-only, interband-only, and total currents. The intraband-only THG is weak (see main text Fig.~\ref{fig:1}e) and has a smaller ellipticity, $\epsilon_{\mathrm{intra}} = 0.36$. In contrast, the interband contribution dominates the harmonic intensity and exhibits a much larger, near-circular ellipticity, $\epsilon_{\mathrm{inter}} = 0.94$. The total THG has $\epsilon_{\mathrm{total}} = 0.63$. The weak intraband admixture modifies the total polarization ellipse and reduces the ellipticity from 0.94 to 0.63; however, this quantitative modification does not constitute the origin of the chiral polarization. The total chiral response is inherited from the coherent superposition of multiple interband transition channels, rather than generated by intra–interband interference. These results show that the intraband contribution cannot account for the observed chiral harmonic emission, and that the elliptical polarization arises from the coherent superposition of multiple interband transition channels.
}

\subsection*{Supplementary Section S13. Harmonic ellipticity under different incident angle}
{In table S1, we compared ellipticity under $10^{\circ}$ and $4^{\circ}$ incidence for 2HG, THG and 4HG. The data shows that $\epsilon$ changes only slightly, indicating that the occurrence of elliptical polarization does not depend on a specific angle of incidence.}
\begin{table}[htbp]
    \centering
    \caption{Harmonic ellipticity under different incident angle}
    \label{tab:my_table}
    \renewcommand{\arraystretch}{1.5} 
    \setlength{\tabcolsep}{10pt} 
    \begin{tabular}{|c|c|c|c|}
        \hline
        & 
        \makecell{(101) 2HG \\ $\theta = 130^{\circ}$} & 
        \makecell{(101) THG \\ $\theta = 147^{\circ}$} & 
        \makecell{(101) 4HG \\ $\theta = 138^{\circ}$} \\
        \hline
        \makecell{incident angle $\sim 10^{\circ}$\\ } &
        \makecell{0.27} & 
        \makecell{0.60} & 
        \makecell{0.65} \\
        \hline
        \makecell{incident angle $\sim 4^{\circ}$\\ } &
        \makecell{0.25} & 
        \makecell{0.50} & 
        \makecell{0.63} \\
        \hline
    \end{tabular}
\end{table}

\subsection*{Supplementary Section S14. dephasing time dependence of polarization state}
\begin{figure}[htbp]
    \begin{center}
	   \includegraphics[page=11, width=1\textwidth, trim=1in 1.7in 1in 1.2in, clip]{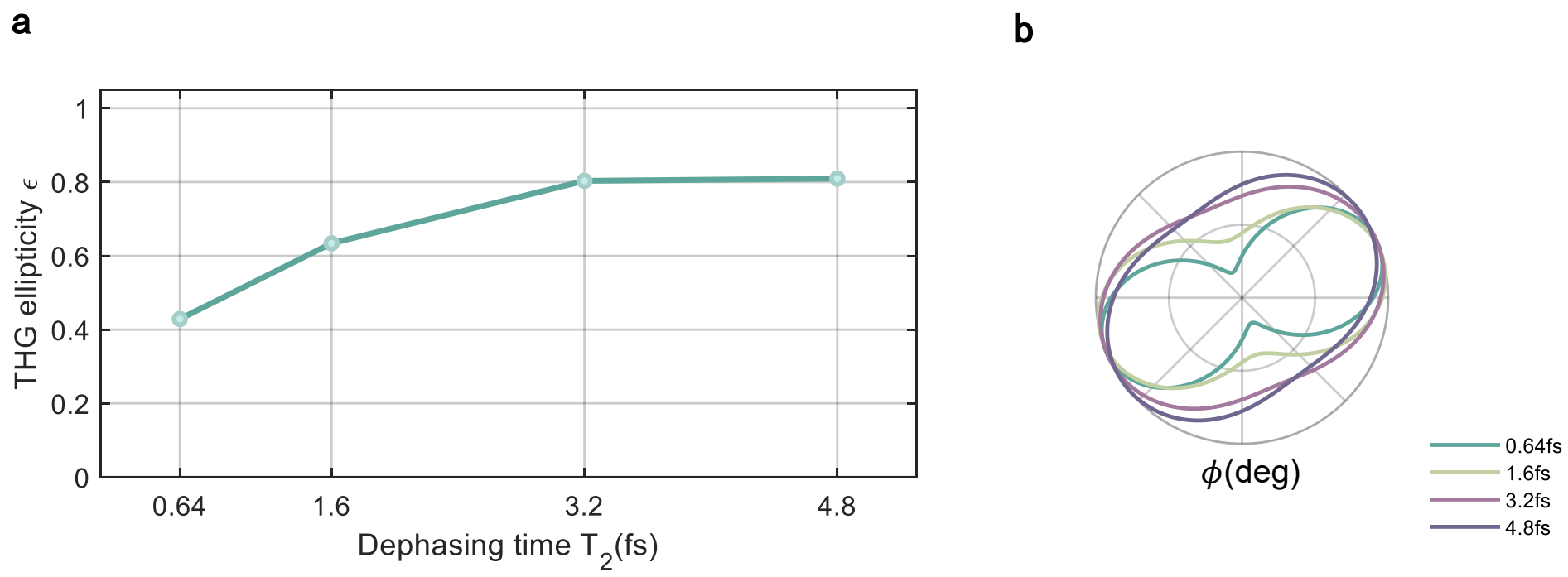}
	   \caption{\textbf{Calculated polarization state dependent on dephasing time.} \textbf{(a)}, Calculated ellipticity of THG for the (101) surface at $\theta = 15^{\circ}$ as a function of the dephasing time $T_2$. The ellipticity varies from 0.43 to 0.81. \textbf{(b)} Polarization state of THG for the (101) surface at $\theta = 15^{\circ}$ as a function of the dephasing time $T_2$. The major-axis orientation remains robust.}
	   \label{fig:S11}
    \end{center}
\end{figure}

{In the SBEs, the dephasing time $T_2$ is introduced as a phenomenological damping term in the equation of motion for the interband polarization, accounting for decoherence induced by electron–electron correlations, electron–phonon coupling, and phonon–phonon scattering. A larger $T_2$ corresponds to weaker dephasing, i.e., less pronounced decoherence of the interband polarization, whereas a smaller $T_2$ implies faster decay of the coherent polarization and stronger broadening.
Physically, dephasing breaks the effective time-reversal relation between excitation and recombination by strongly suppressing electron–hole recombination, thereby lifting the constraint that forbids a current component perpendicular to the linear driver. 
We culculated the polarization state of (101) face at $\theta = 15^{\circ}$ while varing the dephasing time from $T_2 = 0.64$ to $4.81$ fs). The major-axis orientation is found to be robust, whereas the ellipticity varies from 0.43 to 0.81.
Fig.~\ref{fig:S11}a shows the ellipticity as a function of $T_2$. With increasing $T_2$, the dephasing effect becomes weaker and the ellipticity exhibits an upward trend. 
The enhancement of ellipticity with increasing $T_2$ suggests that preserving interband coherence facilitates coherent multichannel dynamics underlying elliptical harmonic emission, whereas decoherence processes tend to suppress this effect under the present conditions.
The calculations in the main text employ $T_2 = 1.6$ fs. Even at this value, the ellipticity remains sizable, indicating that the main conclusions are not sensitive to the detailed choice of $T_2$.
}

\subsection*{Supplementary Section S15. Sensitivity}
{Angular sensitivity. The original angular scan was performed with a $5^{\circ}$ step. To locate the peak and the 5\% tolerance angles more precisely, we interpolated these data onto a dense grid with a step of 0.01 rad (about $0.57^{\circ}$). The peak value is defined as the maximum on this dense grid, without any peak-shape fitting, giving $\epsilon_{max} = 0.86$ at $\theta_p = 158^{\circ}$. To locate the 5\% tolerance angles, we find where the 0.95 $\epsilon_{max}$ level crosses the interpolated angular curve, using linear interpolation between adjacent dense-grid points; this gives $\Delta\theta_{5\%}^{(-)} = 3.3^{\circ}$ and $\Delta\theta_{5\%}^{(+)} = 3.4^{\circ}$. Because the peak is asymmetric, we quote the two sides separately and take the conservative alignment requirement $\sigma_{\theta}$,align $\lesssim 3^{\circ}$. 
}

{Pump-intensity sensitivity. Because higher pump intensities damaged the sample, no data were measured above the highest safely accessible intensity $I_0=2.7\times10^{11}\,\mathrm{W\,cm^{-2}}$. We denote by $\epsilon_0$ the ellipticity measured at $I_0$ and fixed $\theta=158^{\circ}$. We normalize the pump intensity as $I_n=I/I_0$. We define $\tilde S_I=\partial\ln\epsilon/\partial\ln I|_{I_0}$ and the downward 5\% intensity tolerance by $\epsilon(I_0-\Delta I_{5\%}^{(-)})=0.95\epsilon_0$. At $I_0$, $\epsilon_0=0.84$. Direct interpolation of the raw data, which does not rely on a fit model, gives the main result $\Delta I_{5\%}^{(-)}/I_0=5.41\%$. A local linear fit to the four highest-intensity points ($I_n = 0.640 - 1.000$, N = 4) gives $\tilde S_I=1.264\pm0.121$ and $\Delta I_{5\%}^{(-)}/I_0=3.96\%\pm0.38\%$, while a local quadratic fit gives $\tilde S_I=0.699$ and $\Delta I_{5\%}^{(-)}/I_0=6.24\%$. We therefore quote $\Delta I_{5\%}^{(-)}/I_0=5.4\%$ with a model systematic range of 4.0\%–6.2\%. In practice, the pump intensity should be stable to about $\pm2.5\%$ to keep $\epsilon$ within 5\% of $\epsilon_0$.
}
\begin{figure}[htbp]
    \begin{center}
	   \includegraphics[page=12, width=1\textwidth, trim=1in 0.1in 1in 1in, clip]{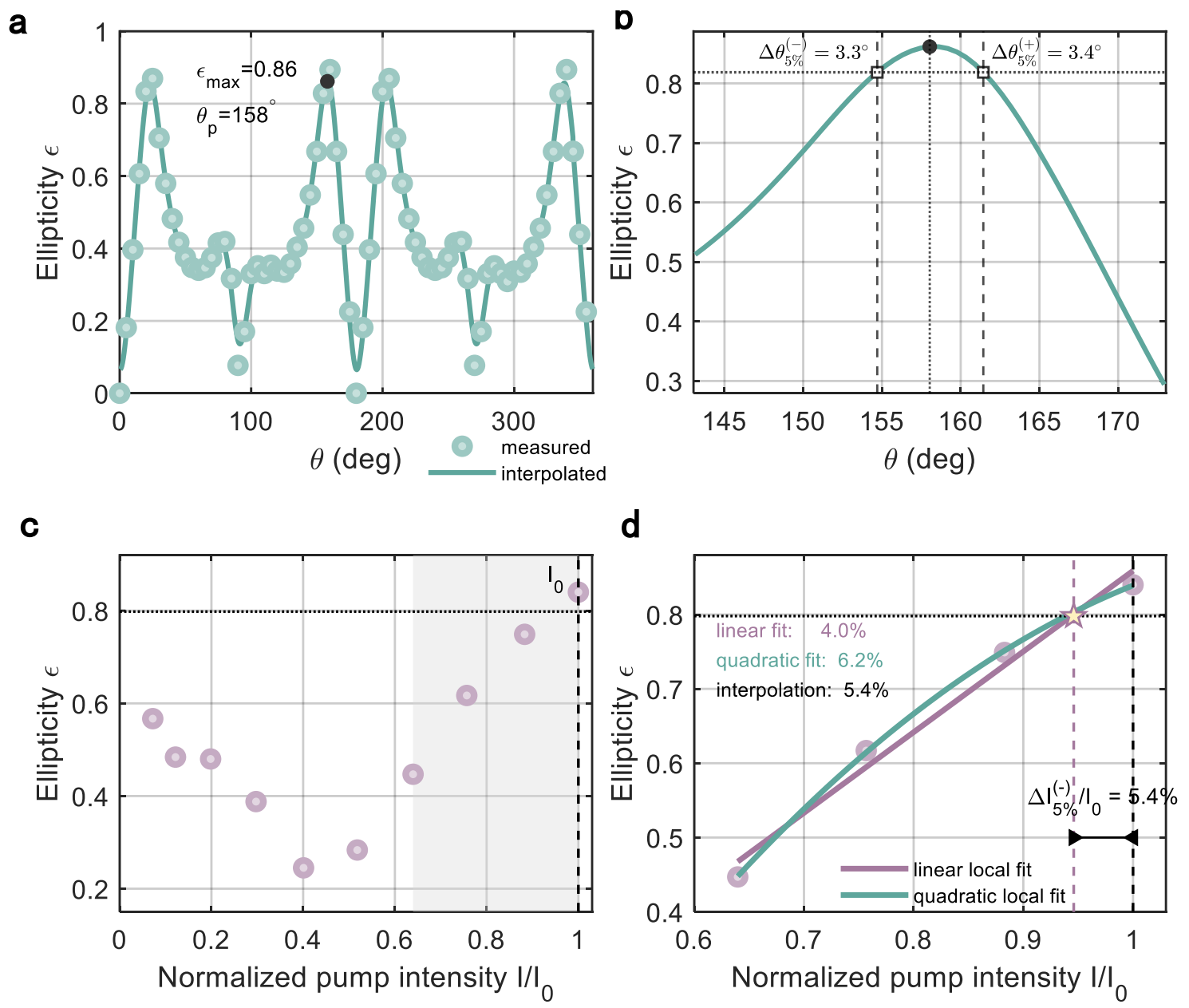}
	   \caption{\textbf{Angular sensitivity and pump-intensity sensitivity of the peak THG ellipticity.} \textbf{(a)(b)}Angular sensitivity of the peak THG ellipticity. \textbf{(a)} Measured THG ellipticity versus driving-polarization angle $\theta$. The light-purple curve is the interpolated experimental data; the dark dot marks the grid maximum, $\epsilon_{max} = 0.862$ at $\theta_p = 158^{\circ}$.
\textbf{(b)} Zoom near the maximum. The black dotted line marks 0.95 $\epsilon_{max}$. The black dashed verticals mark the two crossing angles $\theta_p - \Delta\theta_{5\%}^{(-)}$ and $\theta_p + \Delta\theta_{5\%}^{(+)}$, with $\Delta\theta_{5\%}^{(-)} = 3.3^{\circ}$ and $\Delta\theta_{5\%}^{(+)} = 3.4^{\circ}$. The conservative angular alignment requirement is $\sigma_{\theta}$, align $\lesssim 3^{\circ}$.
\textbf{(c)(d)} Pump-intensity sensitivity of THG ellipticity at $\theta=158^{\circ}$. \textbf{(c)} Ellipticity versus normalized pump intensity $I/I_0$, where $I_0 = 2.7\times10^{11}\,\mathrm{W\,cm^{-2}}$ is the highest safely accessible intensity. The shaded region indicates the local-fit window.
\textbf{(d)} Zoom into the highest-intensity region. Symbols are experimental data; solid lines are local linear and quadratic fits. The black dotted line marks 0.95 $\epsilon_0$, and the dashed vertical marks the interpolated intensity $I_t$ at which $\epsilon = 0.95 \epsilon_0$. Direct interpolation of the raw data gives $\Delta I_{5\%}^{(-)}/I_0 = 5.4\%$; local linear and quadratic fits give 4.0\% and 6.2\%, respectively, which we take as the model-systematic range.}
	   \label{fig:S12}
    \end{center}
\end{figure}

\newpage

\subsection*{Supplementary Section S16. Ellipticity versus number of bands}
\begin{figure}[htbp]
    \begin{center}
	   \includegraphics[page=13, width=1\textwidth, trim=1in 1in 1in 1in, clip]{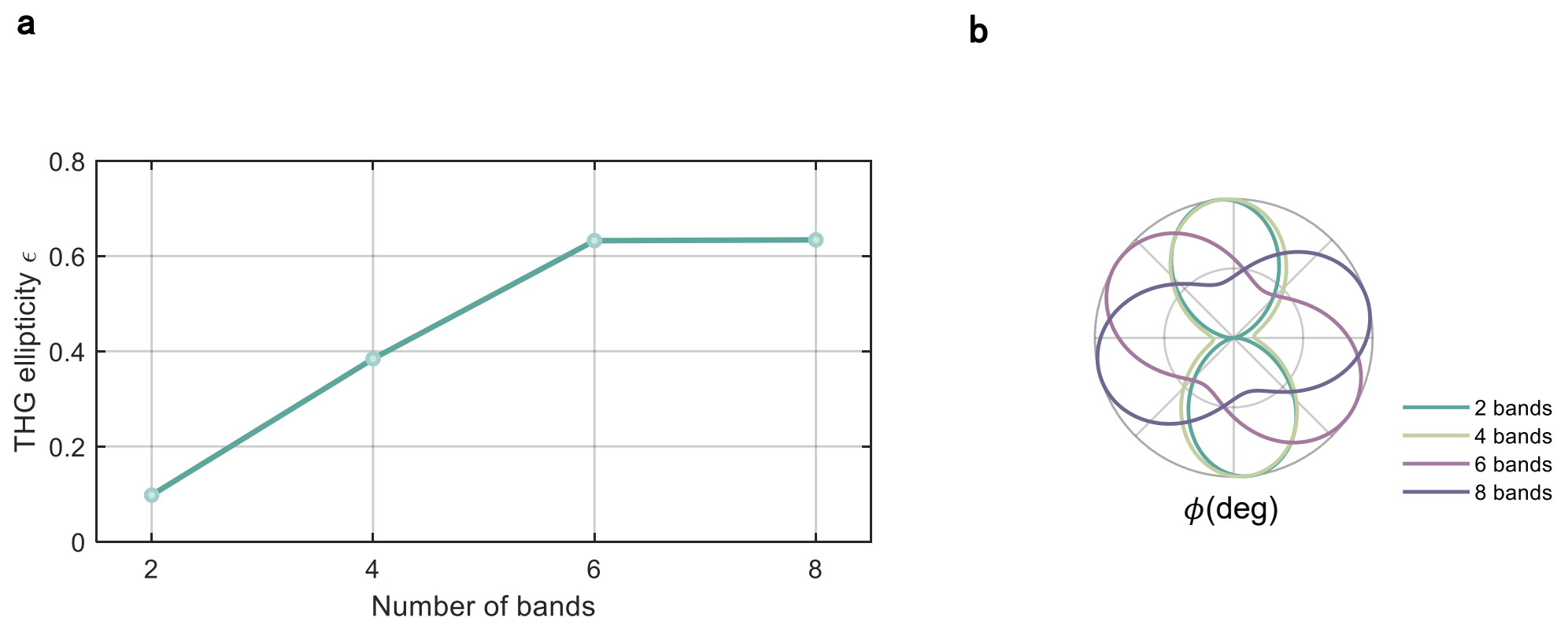}
	   \caption{\textbf{Calculated polarization state versus number of bands.} \textbf{(a)}, Calculated ellipticity of THG for the (101) surface at $\theta = 15^{\circ}$ as a function of number of bands. \textbf{(b)}, Polarization state of THG for the (101) surface at $\theta = 15^{\circ}$ for different number of bands.}
	   \label{fig:S13}
    \end{center}
\end{figure}

{To check the sensitivity of the calculated ellipticity to the band truncation, we performed SBEs with two, four, six, and eight bands under identical driving conditions (1,920 $\mathrm{nm}$, 
$\theta = 15^{\circ}$, peak intensity $\sim 2.7\times10^{11}\,\mathrm{W\,cm^{-2}}$), see Fig.~\ref{fig:S13}. In each case, the total THG ellipticity was extracted from the total current including both interband and intraband contributions. The ellipticity increases with the number of bands and converges by six bands, with the six-band value already close to the full eight-band result (Supplementary Fig. S13). The two-band calculation gives only weak ellipticity, consistent with the symmetry analysis in Supplementary Section S3. The convergence by six bands shows that the calculated ellipticity is not an artifact of including additional high-energy bands. This band-count calculation establishes the inadequacy of the two-band description and the convergence of the result; the direct evidence for multichannel interference is provided by the channel-cumulative analysis in the main text (Fig.~\ref{fig:3}a).
}

\subsection*{Supplementary Section S17.  Intensity-dependent THG yield and channel-resolved cumulative polarization power at $\theta = 15^{\circ}$ and $45^{\circ}$.}
{As shown in Supplementary Fig.~\ref{fig:S14}a, the THG yield at $\theta = 15^{\circ}$ exhibits a pronounced kink, in contrast to the smoother nonperturbative response at $\theta = 45^{\circ}$. To elucidate this orientation dependence, we performed calculations in which individual interband contributions were coherently added in descending order of their THG intensities (Supplementary Fig.~\ref{fig:S14}b,c). At $\theta = 15^{\circ}$, the cumulative major-axis power exhibits a pronounced dip, indicating destructive interference between interband channels, whereas at $\theta = 45^{\circ}$ it increases overall without comparable suppression. Meanwhile, the minor-axis powers at the two orientations remain of the same order of magnitude. This selective suppression of the major-axis contribution enhances the ellipticity at $\theta = 15^{\circ}$. These results, together with Fig.~\ref{fig:2}, establish a physical connection between the angular anisotropy and harmonic ellipticity. Nevertheless, the present calculations do not establish two distinct microscopic processes dominating on either side of the intensity crossover. Accordingly, we now present the angular-pattern evolution and the intensity-dependent kink as supporting phenomenology consistent with orientation-dependent multichannel interband interference, rather than as independent evidence for a specific crossover mechanism.
}

\begin{figure}[htbp]
    \begin{center}
	   \includegraphics[page=14, width=1\textwidth, trim=1in 0.5in 1in 1in, clip]{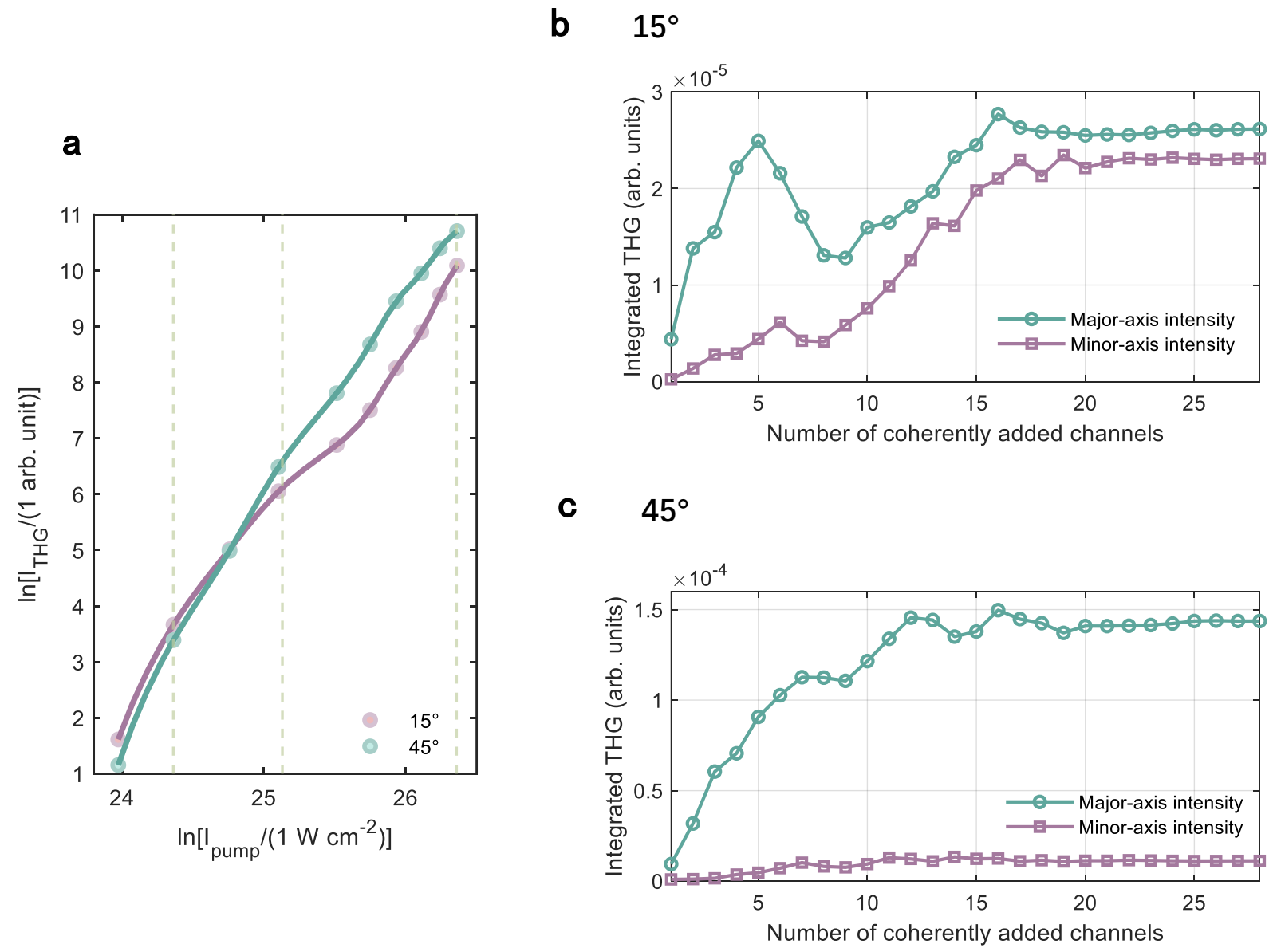}
	   \caption{\textbf{Intensity-dependent THG yield and channel-resolved cumulative polarization power at $\theta = 15^{\circ}$ and $45^{\circ}$.}\textbf{(a)} Pump-intensity dependence of the THG yield at $\theta = 15^{\circ}$ and $45^{\circ}$, showing a pronounced kink at $\theta = 15^{\circ}$ and a smoother nonperturbative dependence at $45^{\circ}$. Purple and green circles are experimental data of $\theta = 15^{\circ}$ and $45^{\circ}$ respectively, while lines are generated by interpolation solely as a visual guide for the eye. \textbf{(b)(c)} SBEs calculated cumulative major-axis THG intensity (green circles) and minor-axis THG intensity (purple squares) obtained by coherently adding interband contributions in descending order of their individual THG intensities at $\theta = 15^{\circ}$ and  $45^{\circ}$.}
	   \label{fig:S14}
    \end{center}
\end{figure}

\newpage
\printbibliography

@Article{MunLight2020,
author={Mun, Jungho
and Kim, Minkyung
and Yang, Younghwan
and Badloe, Trevon
and Ni, Jincheng
and Chen, Yang
and Qiu, Cheng-Wei
and Rho, Junsuk},
title={Electromagnetic chirality: from fundamentals to nontraditional chiroptical phenomena},
journal={Light: Science {\&} Applications},
year={2020},
month={Sep},
day={02},
volume={9},
number={1},
pages={139},
issn={2047-7538},
doi={10.1038/s41377-020-00367-8},
url={https://doi.org/10.1038/s41377-020-00367-8}
}

@Article{HanNature2025,
author={Han, Meng and Ji, Jia-Bao and Blech, Alexander
and Goetz, R. Esteban
and Allison, Corbin
and Greenman, Loren
and Koch, Christiane P.
and W{\"o}rner, Hans Jakob},
title={Attosecond control and measurement of chiral photoionization dynamics},
journal={Nature},
year={2025},
volume={645},
pages={95-100},
url={https://doi.org/10.1038/s41586-025-09455-4}
}

@article{LuoScience2023,
author = {Jiaming Luo  and Tong Lin  and Junjie Zhang  and Xiaotong Chen  and Elizabeth R. Blackert  and Rui Xu  and Boris I. Yakobson  and Hanyu Zhu },
title = {Large effective magnetic fields from chiral phonons in rare-earth halides},
journal = {Science},
volume = {382},
number = {6671},
pages = {698-702},
year = {2023},
doi = {10.1126/science.adi9601},
URL = {https://www.science.org/doi/abs/10.1126/science.adi9601}
}

@article{ChanPRL2016,
  title = {When Chiral Photons Meet Chiral Fermions: Photoinduced Anomalous {Hall} Effects in {Weyl} Semimetals},
  author = {Chan, Ching-Kit and Lee, Patrick A. and Burch, Kenneth S. and Han, Jung Hoon and Ran, Ying},
  journal = {Phys. Rev. Lett.},
  volume = {116},
  issue = {2},
  pages = {026805},
  numpages = {5},
  year = {2016},
  month = {Jan},
  publisher = {American Physical Society},
  doi = {10.1103/PhysRevLett.116.026805},
  url = {https://link.aps.org/doi/10.1103/PhysRevLett.116.026805}
}

@article{DietrichPRA1994,
  title = {High-harmonic generation and correlated two-electron multiphoton ionization with elliptically polarized light},
  author = {Dietrich, P. and Burnett, N. H. and Ivanov, M. and Corkum, P. B.},
  journal = {Phys. Rev. A},
  volume = {50},
  issue = {5},
  pages = {R3585--R3588},
  numpages = {0},
  year = {1994},
  month = {Nov},
  publisher = {American Physical Society},
  doi = {10.1103/PhysRevA.50.R3585},
  url = {https://link.aps.org/doi/10.1103/PhysRevA.50.R3585}
}

@article{DirectIndirectInterfereHHG2015,
  title={Real-time observation of interfering crystal electrons in high-harmonic generation},
  author={Hohenleutner, Matthias and Langer, Fabian and Schubert, Olaf and Knorr, Matthias and Huttner, U and Koch, Stephan W and Kira, M and Huber, Rupert},
  journal={Nature},
  volume={523},
  number={7562},
  pages={572--575},
  year={2015},
  publisher={Nature Publishing Group UK London},
  url = {https://doi.org/10.1038/nature14652}
}

@article{HSGinterferenceBloch2017,
  title={Dynamical birefringence: electron-hole recollisions as probes of Berry curvature},
  author={Banks, Hunter B and Wu, Qile and Valovcin, Darren C and Mack, Shawn and Gossard, Arthur C and Pfeiffer, Loren and Liu, Ren-Bao and Sherwin, Mark S},
  journal={Physical Review X},
  volume={7},
  number={4},
  pages={041042},
  year={2017},
  publisher={APS},
  url = {https://doi.org/10.1103/PhysRevX.7.041042}
}

@article{Linear2ellipticalHHGGaAs2023,
  title={Enhancing high harmonic generation in GaAs by elliptically polarized light excitation},
  author={Sekiguchi, Fumiya and Sakamoto, Minoru and Nakagawa, Kotaro and Tahara, Hirokazu and Sato, Shunsuke A and Hirori, Hideki and Kanemitsu, Yoshihiko},
  journal={Physical Review B},
  volume={108},
  number={20},
  pages={205201},
  year={2023},
  publisher={APS},
  url = {https://link.aps.org/doi/10.1103/PhysRevB.108.205201}
}

@article{MoS2Multiband2022,
  title={Signatures of multiband effects in high-harmonic generation in monolayer {$MoS_2$}},
  author={Yue, Lun and Hollinger, Richard and Uzundal, Can B and Nebgen, Bailey and Gan, Ziyang and Najafidehaghani, Emad and George, Antony and Spielmann, Christian and Kartashov, Daniil and Turchanin, Andrey and others},
  journal={Physical Review Letters},
  volume={129},
  number={14},
  pages={147401},
  year={2022},
  publisher={APS},
  url = {https://doi.org/10.1103/PhysRevLett.129.147401}
}

@article{DynamicSymmInFloquet2017,
  title={Floquet group theory and its application to selection rules in harmonic generation},
  author={Neufeld, Ofer and Podolsky, Daniel and Cohen, Oren},
  journal={Nature Communications},
  volume={10},
  number={1},
  pages={405},
  year={2019},
  publisher={Nature Publishing Group UK London},
  url = {https://doi.org/10.1038/s41467-018-07935-y}
}

@article{qc2024HhgSymm,
  title={Theory of solid-state harmonic generation governed by crystal symmetry},
  author={Qian, Chen and Jiang, Shicheng and Wu, Tong and Weng, Hongming and Yu, Chao and Lu, Ruifeng},
  journal={Physical Review B},
  volume={109},
  number={20},
  pages={205401},
  year={2024},
  publisher={APS}
}

@article{luu2016SBEsQuantumInterference,
  title={High-order harmonic generation in solids: A unifying approach},
  author={Luu, Tran Trung and W{\"o}rner, Hans Jakob},
  journal={Physical Review B},
  volume={94},
  number={11},
  pages={115164},
  year={2016},
  publisher={APS},
  url = {https://doi.org/10.1103/PhysRevB.94.115164}
}

@article{NPho2014CEPstableTHzQuantumInterference,
  title={Sub-cycle control of terahertz high-harmonic generation by dynamical {Bloch} oscillations},
  author={Schubert, Olaf and Hohenleutner, Matthias and Langer, Fabian and Urbanek, Benedikt and Lange, C and Huttner, U and Golde, D and Meier, T and Kira, M and Koch, Stephan W and others},
  journal={Nature Photonics},
  volume={8},
  number={2},
  pages={119--123},
  year={2014},
  publisher={Nature Publishing Group UK London},
  url = {https://doi.org/10.1038/nphoton.2013.349}
}

@article{2015NpNbpMagnetoresistance,
  title={Extremely large magnetoresistance and ultrahigh mobility in the topological {Weyl} semimetal candidate {NbP}},
  author={Shekhar, Chandra and Nayak, Ajaya K and Sun, Yan and Schmidt, Marcus and Nicklas, Michael and Leermakers, Inge and Zeitler, Uli and Skourski, Yurii and Wosnitza, Jochen and Liu, Zhongkai and others},
  journal={Nature Physics},
  volume={11},
  number={8},
  pages={645--649},
  year={2015},
  publisher={Nature Publishing Group UK London},
  url = {https://doi.org/10.1038/nphys3372}
}

@article{2022PrbWeyl2HHGtheory,
  title={Nonlinear optical response of type-{II} {Weyl} fermions in two dimensions},
  author={Tamashevich, Yaraslau and Villari, Leone Di Mauro and Ornigotti, Marco},
  journal={Physical Review B},
  volume={105},
  number={19},
  pages={195102},
  year={2022},
  publisher={APS},
  doi = {10.1103/PhysRevB.105.195102},
  url = {https://link.aps.org/doi/10.1103/PhysRevB.105.195102}
}

@article{2022PraHHGin3dWeylTheory,
  title={High-order harmonic generation in three-dimensional {Weyl} semimetals with broken time-reversal symmetry},
  author={Avetissian, HK and Avetisyan, VN and Avchyan, BR and Mkrtchian, GF},
  journal={Physical Review A},
  volume={106},
  number={3},
  pages={033107},
  year={2022},
  publisher={APS},
  url = {https://doi.org/10.1103/PhysRevA.106.033107}
}

@article{2022PrbWeylAHETheory,
  title={High-harmonic spectroscopy of light-driven nonlinear anisotropic anomalous {Hall} effect in a {Weyl} semimetal},
  author={Bharti, Amar and Mrudul, MS and Dixit, Gopal},
  journal={Physical Review B},
  volume={105},
  number={15},
  pages={155140},
  year={2022},
  publisher={APS},
  doi = {10.1103/PhysRevB.105.155140},
  url = {https://link.aps.org/doi/10.1103/PhysRevB.105.155140}
}

@article{2024PrbWeylMergingWeylNodeHHGTheory,
  title={High-harmonic generation in semi-{Dirac} and {Weyl} semimetals with broken time-reversal symmetry: Exploration of the merging of {Weyl} nodes},
  author={Medic, Luka and Mravlje, Jernej and Ram{\v{s}}ak, Anton and Rejec, Toma{\v{z}}},
  journal={Physical Review B},
  volume={109},
  number={20},
  pages={205130},
  year={2024},
  publisher={APS},
  doi = {10.1103/PhysRevB.109.205130},
  url = {https://link.aps.org/doi/10.1103/PhysRevB.109.205130}
}

@article{2024AplWeylHHGIntensityTuningTheory,
  title={Non-perturbative nonlinear optical responses in {Weyl} semimetals},
  author={Bharti, Amar and Dixit, Gopal},
  journal={Applied Physics Letters},
  volume={125},
  number={5},
  year={2024},
  publisher={AIP Publishing},
  url = {https://doi.org/10.1063/5.0217186}
}

@article{2023PrbRoleOfTopoChargeWeylHHGTheory,
  title={Role of topological charges in the nonlinear optical response from {Weyl} semimetals},
  author={Bharti, Amar and Dixit, Gopal},
  journal={Physical Review B},
  volume={107},
  number={22},
  pages={224308},
  year={2023},
  publisher={APS},
  doi = {10.1103/PhysRevB.107.224308},
  url = {https://link.aps.org/doi/10.1103/PhysRevB.107.224308}
}

@article{lv2021WeylWP2HHG,
  title={High-harmonic generation in {Weyl} semimetal {$\beta$}-{$WP_2$} crystals},
  author={Lv, Yang-Yang and Xu, Jinlong and Han, Shuang and Zhang, Chi and Han, Yadong and Zhou, Jian and Yao, Shu-Hua and Liu, Xiao-Ping and Lu, Ming-Hui and Weng, Hongming and others},
  journal={Nature Communications},
  volume={12},
  number={1},
  pages={6437},
  year={2021},
  publisher={Nature Publishing Group UK London},
  url = {https://doi.org/10.1038/s41467-021-26766-y}
}

@article{wu2017TaAsSHG,
  title={Giant anisotropic nonlinear optical response in transition metal monopnictide {Weyl} semimetals},
  author={Wu, Liang and Patankar, Shreyas and Morimoto, Takahiro and Nair, Nityan L and Thewalt, Eric and Little, Arielle and Analytis, James G and Moore, Joel E and Orenstein, Joseph},
  journal={Nature Physics},
  volume={13},
  number={4},
  pages={350--355},
  year={2017},
  publisher={Nature Publishing Group UK London},
  url = {https://doi.org/10.1038/nphys3969}
}

@article{xu2015TaAsFermiArc,
  title={Discovery of a {Weyl} fermion semimetal and topological Fermi arcs},
  author={Xu, Su-Yang and Belopolski, Ilya and Alidoust, Nasser and Neupane, Madhab and Bian, Guang and Zhang, Chenglong and Sankar, Raman and Chang, Guoqing and Yuan, Zhujun and Lee, Chi-Cheng and others},
  journal={Science},
  volume={349},
  number={6248},
  pages={613--617},
  year={2015},
  publisher={American Association for the Advancement of Science},
  doi = {10.1126/science.aaa9297},
URL = {https://www.science.org/doi/abs/10.1126/science.aaa9297},
}

@article{lv2015TaAsfermiArc,
  title = {Experimental Discovery of {Weyl} Semimetal {TaAs}},
  author = {Lv, B. Q. and Weng, H. M. and Fu, B. B. and Wang, X. P. and Miao, H. and Ma, J. and Richard, P. and Huang, X. C. and Zhao, L. X. and Chen, G. F. and Fang, Z. and Dai, X. and Qian, T. and Ding, H.},
  journal = {Phys. Rev. X},
  volume = {5},
  issue = {3},
  pages = {031013},
  numpages = {8},
  year = {2015},
  month = {Jul},
  publisher = {American Physical Society},
  doi = {10.1103/PhysRevX.5.031013},
  url = {https://link.aps.org/doi/10.1103/PhysRevX.5.031013}
}

@article{lv2015TaAsWeylNode,
  title={Observation of {Weyl} nodes in {TaAs}},
  author={Lv, BQ and Xu, N and Weng, HM and Ma, JZ and Richard, P and Huang, XC and Zhao, LX and Chen, GF and Matt, CE and Bisti, F and others},
  journal={Nature Physics},
  volume={11},
  number={9},
  pages={724--727},
  year={2015},
  publisher={Nature Publishing Group UK London},
  url = {https://doi.org/10.1038/nphys3426}
}

@article{zhang2016TaAsNegativeLMR,
  title={Signatures of the {Adler--Bell--Jackiw} chiral anomaly in a {Weyl} fermion semimetal},
  author={Zhang, Cheng-Long and Xu, Su-Yang and Belopolski, Ilya and Yuan, Zhujun and Lin, Ziquan and Tong, Bingbing and Bian, Guang and Alidoust, Nasser and Lee, Chi-Cheng and Huang, Shin-Ming and others},
  journal={Nature Communications},
  volume={7},
  number={1},
  pages={1--9},
  year={2016},
  publisher={Nature Publishing Group},
  url = {https://doi.org/10.1038/ncomms10735}
}

@article{huang2015TaAsNegativeLMR,
  title={Observation of the chiral-anomaly-induced negative magnetoresistance in {$3D$} {Weyl} semimetal {TaAs}},
  author={Huang, Xiaochun and Zhao, Lingxiao and Long, Yujia and Wang, Peipei and Chen, Dong and Yang, Zhanhai and Liang, Hui and Xue, Mianqi and Weng, Hongming and Fang, Zhong and others},
  journal={Physical Review X},
  volume={5},
  number={3},
  pages={031023},
  year={2015},
  publisher={APS},
  url = {https://doi.org/10.1103/PhysRevX.5.031023}
}

@article{ma2017TaAsPumpCPGE,
  title={Direct optical detection of {Weyl} fermion chirality in a topological semimetal},
  author={Ma, Qiong and Xu, Su-Yang and Chan, Ching-Kit and Zhang, Cheng-Long and Chang, Guoqing and Lin, Yuxuan and Xie, Weiwei and Palacios, Tom{\'a}s and Lin, Hsin and Jia, Shuang and others},
  journal={Nature Physics},
  volume={13},
  number={9},
  pages={842--847},
  year={2017},
  publisher={Nature Publishing Group UK London},
  url = {https://doi.org/10.1038/nphys4146}
}

@article{RanaPRB2024,
  title={High-harmonic generation from strain-engineered graphene for polarization tailoring},
  author={Rana, Navdeep and Mrudul, MS and Dixit, Gopal},
  journal={Physical Review B},
  volume={110},
  number={5},
  pages={054103},
  year={2024},
  publisher={APS},
  url = {https://doi.org/10.1103/PhysRevB.110.054103}
}

@article{Dong2021TheoryLinear2EllipticalGraphene,
Author = {Dong, Fulong and Xia, Qinzhi and Liu, Jie},
Title = {Ellipticity of the harmonic emission from graphene irradiated by a linearly polarized laser},
Journal = {Physical Review A},
Year = {2021},
Volume = {104},
pages = {033119},
Number = {3},
Month = {SEP 30},
DOI = {10.1103/PhysRevA.104.033119},
Article-Number = {033119},
ISSN = {2469-9926},
EISSN = {2469-9934},
ResearcherID-Numbers = {Liu, Jie/D-2041-2017},
ORCID-Numbers = {Dong, Fulong/0000-0001-9210-295X
   },
Unique-ID = {WOS:000704065100007},
url = {https://link.aps.org/doi/10.1103/PhysRevA.104.033119}
}

@article{Zurron2019TheoryLinear2EllipticalGraphene,
Author = {Zurron-Cifuentes, Oscar and Boyero-Garcia, Roberto and Hernandez-Garcia,
   Carlos and Picon, Antonio and Plaja, Luis},
Title = {Optical anisotropy of non-perturbative high-order harmonic generation in
   gapless graphene},
Journal = {Optics Express},
Year = {2019},
Volume = {27},
Number = {5},
Pages = {7776-7786},
Month = {MAR 4},
DOI = {10.1364/OE.27.007776},
ISSN = {1094-4087},
ResearcherID-Numbers = {ZURRÓN-CIFUENTES, ÓSCAR/AAC-3426-2021
   HERNANDEZ GARCIA, CARLOS/G-3681-2011
   Hernandez-Garcia, Carlos/G-3681-2011
   Boyero-García, Roberto/AFR-1286-2022
   Picon Alvarez, Antonio/I-7268-2012
   Plaja, Luis/K-8701-2014
   Picón, Antonio/I-7268-2012},
ORCID-Numbers = {HERNANDEZ GARCIA, CARLOS/0000-0002-6153-2647
   Boyero-Garcia, Roberto/0000-0002-9475-9442
   Picon Alvarez, Antonio/0000-0002-6142-3440
   ZURRON-CIFUENTES, OSCAR/0000-0002-4510-8264
   },
Unique-ID = {WOS:000460170000150},
}

@article{kfir2015circularEUV,
  title={Generation of bright phase-matched circularly-polarized extreme ultraviolet high harmonics},
  author={Kfir, Ofer and Grychtol, Patrik and Turgut, Emrah and Knut, Ronny and Zusin, Dmitriy and Popmintchev, Dimitar and Popmintchev, Tenio and Nembach, Hans and Shaw, Justin M and Fleischer, Avner and others},
  journal={Nature Photonics},
  volume={9},
  number={2},
  pages={99--105},
  year={2015},
  publisher={Nature Publishing Group UK London},
  URL = {https://doi.org/10.1038/nphoton.2014.293}
}

@article{wang2024ChiralPhononReview,
  title={Chiral phonons: Prediction, verification, and application},
  author={Wang, Tingting and Sun, Hong and Li, Xiaozhe and Zhang, Lifa},
  journal={Nano Letters},
  volume={24},
  number={15},
  pages={4311--4318},
  year={2024},
  publisher={ACS Publications},
  url = {https://doi.org/10.1021/acs.nanolett.4c00606}
}

@article{ Fan2015GenerationEUV,
Author = {Fan, Tingting and Grychtol, Patrik and Knut, Ronny and Hernandez-Garcia,
   Carlos and Hickstein, Daniel D. and Zusin, Dmitriy and Gentry, Christian
   and Dollar, Franklin J. and Mancuso, Christopher A. and Hogle, Craig W.
   and Kfir, Ofer and Legut, Dominik and Carva, Karel and Ellis, Jennifer
   L. and Dorney, Kevin M. and Chen, Cong and Shpyrko, Oleg G. and
   Fullerton, Eric E. and Cohen, Oren and Oppeneer, Peter M. and Milosevic,
   Dejan B. and Becker, Andreas and Jaron-Becker, Agnieszka A. and
   Popmintchev, Tenio and Murnane, Margaret M. and Kapteyn, Henry C.},
Title = {Bright circularly polarized soft {X}-ray high harmonics for {X}-ray magnetic
   circular dichroism},
Journal = {Proceedings of the National Academy of Sciences of the United States of America},
Year = {2015},
Volume = {112},
Number = {46},
Pages = {14206-14211},
Month = {NOV 17},
DOI = {10.1073/pnas.1519666112},
ISSN = {0027-8424},
}

@article{ Skantzakis2016GenerationEUVthroughPreAlinedMolecule,
Author = {Skantzakis, E. and Chatziathanasiou, S. and Carpeggiani, P. A. and
   Sansone, G. and Nayak, A. and Gray, D. and Tzallas, P. and
   Charalambidis, D. and Hertz, E. and Faucher, O.},
Title = {Polarization shaping of high-order harmonics in laser-aligned molecules},
Journal = {Scientific Reports},
Year = {2016},
Volume = {6},
Pages = {39295},
Month = {DEC 20},
Article-Number = {39295},
ISSN = {2045-2322},
url = {https://doi.org/10.1038/srep39295},
}

@article{ Kim2021Spintronics,
Author = {Kim, Young-Hoon and Zhai, Yaxin and Lu, Haipeng and Pan, Xin and Xiao,
   Chuanxiao and Gaulding, E. Ashley and Harvey, Steven P. and Berry,
   Joseph J. and Vardeny, Zeev Valy and Luther, Joseph M. and Beard,
   Matthew C.},
Title = {Chiral-induced spin selectivity enables a room-temperature spin
   light-emitting diode},
Journal = {Science},
Year = {2021},
Volume = {371},
Number = {6534},
Pages = {1129},
Month = {MAR 12},
DOI = {10.1126/science.abf5291},
ISSN = {0036-8075},
EISSN = {1095-9203},
Unique-ID = {WOS:000630096400028},
}

@article{Ma2025prlVdwPolarizationControl,
  title = {Phase Matching of High Harmonic Generation in Twisted van der {Waals} Crystals},
  author = {Ma, Chenjun and Huang, Chen and You, Yilong and Liu, Huazhan and Ding, Zhitong and Ding, Mingchao and Liu, Chang and Zhang, Jin and Li, Guixin and Sun, Zhipei and Wu, Shiwei and Ma, Chaojie and Wang, Enge and Hong, Hao and Liu, Kaihui},
  journal = {Phys. Rev. Lett.},
  volume = {135},
  issue = {16},
  pages = {166902},
  numpages = {6},
  year = {2025},
  month = {Oct},
  publisher = {American Physical Society},
  doi = {10.1103/kskc-qlb3},
  url = {https://link.aps.org/doi/10.1103/kskc-qlb3}
}

@article{PhysRevA.49.2117,
  title = {Theory of high-harmonic generation by low-frequency laser fields},
  author = {Lewenstein, M. and Balcou, Ph. and Ivanov, M. Yu. and L'Huillier, Anne and Corkum, P. B.},
  journal = {Phys. Rev. A},
  volume = {49},
  issue = {3},
  pages = {2117--2132},
  numpages = {0},
  year = {1994},
  month = {Mar},
  publisher = {American Physical Society},
  doi = {10.1103/PhysRevA.49.2117},
  url = {https://link.aps.org/doi/10.1103/PhysRevA.49.2117}
}

@article{PrbWP2Calculation,
  title = {Theoretical investigations of high harmonic generation in the {Weyl} semimetal {$WP_2$} },
  author = {Liu, Xiulan and Geng, Lei and Kong, Xiao-Shuang and Zhang, Jianing and Fang, Yong-Kang and Peng, Liang-You},
  journal = {Phys. Rev. B},
  volume = {111},
  issue = {18},
  pages = {184314},
  numpages = {12},
  year = {2025},
  month = {May},
  publisher = {American Physical Society},
  doi = {10.1103/PhysRevB.111.184314},
  url = {https://link.aps.org/doi/10.1103/PhysRevB.111.184314}
}

@article{2022SekiguchiPrbGaAsLinear2Elliptical,
  title = {Polarization anomaly in high harmonics in the crossover region between perturbative and extreme nonlinearity in GaAs},
  author = {Sekiguchi, Fumiya and Yumoto, Go and Hirori, Hideki and Kanemitsu, Yoshihiko},
  journal = {Phys. Rev. B},
  volume = {106},
  issue = {24},
  pages = {L241201},
  numpages = {6},
  year = {2022},
  month = {Dec},
  publisher = {American Physical Society},
  doi = {10.1103/PhysRevB.106.L241201},
  url = {https://link.aps.org/doi/10.1103/PhysRevB.106.L241201}
}

@article{2009CO2linear2ellipCalculation,
  title = {Attosecond Circular Dichroism Spectroscopy of Polyatomic Molecules},
  author = {Smirnova, Olga and Patchkovskii, Serguei and Mairesse, Yann and Dudovich, Nirit and Villeneuve, David and Corkum, Paul and Ivanov, Misha Yu.},
  journal = {Phys. Rev. Lett.},
  volume = {102},
  issue = {6},
  pages = {063601},
  numpages = {4},
  year = {2009},
  month = {Feb},
  publisher = {American Physical Society},
  doi = {10.1103/PhysRevLett.102.063601},
  url = {https://link.aps.org/doi/10.1103/PhysRevLett.102.063601}
}

@article{2010PrlN2linear2ellipExp,
  title = {High Harmonic Spectroscopy of Multichannel Dynamics in Strong-Field Ionization},
  author = {Mairesse, Y. and Higuet, J. and Dudovich, N. and Shafir, D. and Fabre, B. and M\'evel, E. and Constant, E. and Patchkovskii, S. and Walters, Z. and Ivanov, M. Yu. and Smirnova, O.},
  journal = {Phys. Rev. Lett.},
  volume = {104},
  issue = {21},
  pages = {213601},
  numpages = {4},
  year = {2010},
  month = {May},
  publisher = {American Physical Society},
  doi = {10.1103/PhysRevLett.104.213601},
  url = {https://link.aps.org/doi/10.1103/PhysRevLett.104.213601}
}

@article{2024PRAhBNparaSHG,
  title = {Tunneling displacement of electron-hole pairs in high-order harmonic generation for monolayer hexagonal boron nitride},
  author = {Dong, Fulong and Yan, Jiayu and Song, Xinru},
  journal = {Phys. Rev. A},
  volume = {110},
  issue = {4},
  pages = {043104},
  numpages = {11},
  year = {2024},
  month = {Oct},
  publisher = {American Physical Society},
  doi = {10.1103/PhysRevA.110.043104},
  url = {https://link.aps.org/doi/10.1103/PhysRevA.110.043104}
}

@article{Kim2025Wp2Kinks,
  title={Quantum interference and occupation control in high harmonic generation from monolayer WS2},
  author={Kim, Minjeong and Kim, Taeho and Galler, Anna and Kim, Dasol and Chacon, Alexis and Gong, Xiangxin and Yang, Yuhui and Fang, Rouli and Watanabe, Kenji and Taniguchi, Takashi and others},
  journal={Nature Communications},
  volume={16},
  number={1},
  pages={9825},
  year={2025},
  publisher={Nature Publishing Group UK London}
}

@article{silva2019topological,
  title={Topological strong-field physics on sub-laser-cycle timescale},
  author={Silva, REF and Jim{\'e}nez-Gal{\'a}n, {\'A} and Amorim, B and Smirnova, O and Ivanov, M},
  journal={Nature Photonics},
  volume={13},
  number={12},
  pages={849--854},
  year={2019},
  publisher={Nature Publishing Group UK London},
  url = {https://doi.org/10.1038/s41566-019-0516-1}
}

@article{tilmann2022NbPFilm3HG,
  title={Ultrafast Sub-100 fs All-Optical Modulation and Efficient Third-Harmonic Generation in {Weyl} Semimetal Niobium Phosphide Thin Films},
  author={Tilmann, Benjamin and Pandeya, Avanindra Kumar and Grinblat, Gustavo and Menezes, Leonardo de S and Li, Yi and Shekhar, Chandra and Felser, Claudia and Parkin, Stuart SP and Bedoya-Pinto, Amilcar and Maier, Stefan A},
  journal={Advanced Materials},
  volume={34},
  number={15},
  pages={2106733},
  year={2022},
  publisher={Wiley Online Library},
  url = {https://doi.org/10.1002/adma.202106733}
}

@article{osterhoudt2019colossal,
  title={Colossal mid-infrared bulk photovoltaic effect in a type-{I} {Weyl} semimetal},
  author={Osterhoudt, Gavin B and Diebel, Laura K and Gray, Mason J and Yang, Xu and Stanco, John and Huang, Xiangwei and Shen, Bing and Ni, Ni and Moll, Philip JW and Ran, Ying and others},
  journal={Nature Materials},
  volume={18},
  number={5},
  pages={471--475},
  year={2019},
  publisher={Nature Publishing Group UK London},
  URL = {https://doi.org/10.1038/s41563-019-0297-4}
}

@article{ma2019nonlinear,
  title={Nonlinear photoresponse of type-{II} {Weyl} semimetals},
  author={Ma, Junchao and Gu, Qiangqiang and Liu, Yinan and Lai, Jiawei and Yu, Peng and Zhuo, Xiao and Liu, Zheng and Chen, Jian-Hao and Feng, Ji and Sun, Dong},
  journal={Nature Materials},
  volume={18},
  number={5},
  pages={476--481},
  year={2019},
  publisher={Nature Publishing Group UK London},
  url = {https://doi.org/10.1038/s41563-019-0296-5}
}

@article{Wilhelm2021,
  author  = {J. Wilhelm and P. Gr{\"o}ssing and A. Seith and J. Crewse and M. Nitsch and L. Weigl and C. Schmid and F. Evers},
  title   = {Semiconductor {Bloch}-equations formalism: Derivation and Application to High-Harmonic Generation from {Dirac} Fermions},
  journal = {Phys. Rev. B},
  volume  = {103},
  pages   = {125419},
  year    = {2021},
  doi = {10.1103/PhysRevB.103.125419},
  url = {https://link.aps.org/doi/10.1103/PhysRevB.103.125419}
}

@article{Golde2008,
  author  = {D. Golde and T. Meier and S. W. Koch},
  title   = {High Harmonics Generated in Semiconductor Nanostructures by the Coupled Dynamics of Optical Inter- and Intraband Excitations},
  journal = {Phys. Rev. B},
  volume  = {77},
  pages   = {075330},
  year    = {2008},
  doi = {10.1103/PhysRevB.77.075330},
  url = {https://link.aps.org/doi/10.1103/PhysRevB.77.075330}

}

@article{Vampa2014,
  author  = {G. Vampa and C. R. McDonald and G. Orlando and D. D. Klug and P. B. Corkum and T. Brabec},
  title   = {Theoretical Analysis of High-Harmonic Generation in Solids},
  journal = {Phys. Rev. Lett.},
  volume  = {113},
  pages   = {073901},
  year    = {2014},
  doi = {10.1103/PhysRevLett.113.073901},
  url = {https://link.aps.org/doi/10.1103/PhysRevLett.113.073901}
}

@article{Li2019,
  author  = {J. Li and X. Zhang and S. Fu and Y. Feng and B. Hu and H. Du},
  title   = {Phase Invariance of the Semiconductor {Bloch} Equations},
  journal = {Phys. Rev. A},
  volume  = {100},
  pages   = {043404},
  year    = {2019},
  doi = {10.1103/PhysRevA.100.043404},
  url = {https://link.aps.org/doi/10.1103/PhysRevA.100.043404}
}

@article{Yue2020,
  author  = {L. Yue and M. B. Gaarde},
  title   = {Structure Gauges and Laser Gauges for the Semiconductor {Bloch} Equations in High-Order Harmonic Generation in Solids},
  journal = {Phys. Rev. A},
  volume  = {101},
  pages   = {053411},
  year    = {2020},
  doi = {10.1103/PhysRevA.101.053411},
  url = {https://link.aps.org/doi/10.1103/PhysRevA.101.053411}
}

@article{VASP,
  author  = {G. Kresse and J. Furthm{\"u}ller},
  title   = {Efficiency of Ab-Initio Total Energy Calculations for Metals and Semiconductors Using a Plane-Wave Basis Set},
  journal = {Comput. Mater. Sci.},
  volume  = {6},
  pages   = {15},
  year    = {1996},
  doi = {10.1103/PhysRevB.54.11169},
  url = {https://link.aps.org/doi/10.1103/PhysRevB.54.11169}
}

@article{Wannier90,
  author  = {A. A. Mostofi and J. R. Yates and Y. S. Lee and I. Souza and D. Vanderbilt and N. Marzari},
  title   = {Wannier90: A Tool for Obtaining Maximally-Localized Wannier Functions},
  journal = {Comput. Phys. Commun.},
  volume  = {178},
  pages   = {685},
  year    = {2008}
}

@article{WannierTools,
  author  = {Q. Wu and S. Zhang and H. F. Song and M. Troyer and A. A. Soluyanov},
  title   = {WannierTools: An Open-Source Software Package for Novel Topological Materials},
  journal = {Comput. Phys. Commun.},
  volume  = {224},
  pages   = {405},
  year    = {2018}
}

@article{Silva2019,
  author  = {R. E. F. Silva and F. Mart{\'i}n and M. Ivanov},
  title   = {High Harmonic Generation in Crystals Using Maximally Localized Wannier Functions},
  journal = {Phys. Rev. B},
  volume  = {100},
  pages   = {195201},
  year    = {2019},
  doi = {10.1103/PhysRevB.100.195201},
  url = {https://link.aps.org/doi/10.1103/PhysRevB.100.195201}
}

@article{PhysRevB.98.195203,
  title = {Optical conductivity of the Weyl semimetal NbP},
  author = {Neubauer, David and Yaresko, Alexander and Li, Weiwu and L\"ohle, Anja and H\"ubner, Ralph and Schilling, Micha B. and Shekhar, Chandra and Felser, Claudia and Dressel, Martin and Pronin, Artem V.},
  journal = {Phys. Rev. B},
  volume = {98},
  issue = {19},
  pages = {195203},
  numpages = {7},
  year = {2018},
  month = {Nov},
  publisher = {American Physical Society},
  doi = {10.1103/PhysRevB.98.195203},
  url = {https://link.aps.org/doi/10.1103/PhysRevB.98.195203}
}

@article{Lai2025,
  author  = {Lai, Fangxing and Yin, Jun and Toftul, Ivan and Barkaoui, Hamdi and Deng, Huachun and Sha, Xinbo and Gorkunov, Maxim V. and Kivshar, Yuri and Song, Qinghai},
  title   = {Nonlinear chiral light generation from resonant metasurfaces},
  journal = {Nature Communications},
  year    = {2025},
  volume  = {16},
  number  = {1},
  pages   = {10686},
  doi     = {10.1038/s41467-025-65721-z},
  url     = {https://doi.org/10.1038/s41467-025-65721-z}
}

\end{document}